\documentclass[preprint]{aastex6}
\usepackage{graphicx}
\usepackage{epsfig}
\usepackage{multirow}
\usepackage{fancyhdr}
\usepackage{footnote}
\usepackage{url}
\usepackage{tablefootnote}

\fancypagestyle{gaiacolor}{\fancyhf{}\fancyfoot[C]{$^4$ \scriptsize{\url{https://gea.esac.esa.int/archive/documentation/GDR2/Data_processing/chap_cu5pho/sec_cu5pho_calibr/ssec_cu5pho_PhotTransf.html}}}}

\shorttitle{Dusty Old Dipper}
\shortauthors{C. Melis {\it et al.}}

\begin{document}


\title{\large \bf Highly structured inner planetary system debris around the intermediate age Sun-like star TYC\,8830~410~1}


\author{\large Carl Melis}
\affil{email: cmelis@ucsd.edu \\
Center for Astrophysics and Space Sciences, University of California, San Diego, CA 92093-0424, USA} 

\author{\large Johan Olofsson}
\affil{Instituto de F\'isica y Astronom\'ia, Facultad de Ciencias, Universidad de Valpara\'iso, Av. Gran Breta\~na 1111, Playa Ancha, Valpara\'iso, Chile\\
and
N\'ucleo Milenio de Formaci\'on Planetaria (NPF), Chile\\}

\author{\large Inseok Song}
\affil{Department of Physics and Astronomy, University of Georgia, Athens, GA 30602-2451, USA}

\author{\large Paula Sarkis}
\affil{Max-Planck-Institut f\"ur Astronomie, K\"onigstuhl 17, Heidelberg 69117, Germany}

\author{\large Alycia J.\ Weinberger}
\affil{Earth and Planets Laboratory, Carnegie Institution for Science, 5241 Broad Branch Rd NW, Washington, DC 20015, USA}

\author{\large Grant Kennedy}
\affil{Department of Physics, University of Warwick, Gibbet Hill Road, Coventry CV4 7AL, UK}

\and

\author{\large Mirko Krumpe}
\affil{Leibniz-Institut f\"{u}r Astrophysik Potsdam (AIP), An der Sternwarte 16, 14482 Potsdam, Germany}









\begin{abstract}

\large{
We present detailed characterization of the extremely dusty main sequence star
TYC\,8830~410~1. 
This system hosts inner planetary system dust (T$_{\rm dust}$$\approx$300\,K)
with a fractional infrared luminosity of $\sim$1\%.
Mid-infrared spectroscopy reveals a strong, mildly-crystalline solid-state emission feature.
TYC\,8830~410~1 (spectral type G9\,V) has a 49.5$'$$'$ separation M4-type companion
co-moving and co-distant with it, and we estimate 
a system age of  $\sim$600\,Myr.
TYC\,8830~410~1 also experiences ``dipper''-like dimming events
as detected by ASAS-SN, {\it TESS}, and characterized in more
detail with the LCOGT.
These recurring eclipses suggest at least one roughly star-sized cloud of dust orbits the star
in addition to assorted smaller dust structures.
The extreme properties of the material orbiting TYC\,8830~410~1 point to dramatic dust-production mechanisms that likely included something similar to the giant-impact event thought to have formed the Earth-Moon system, although hundreds of millions of years after such processes are thought to have concluded in the solar system.
TYC\,8830~410~1 holds promise to deliver significant advances in our understanding of the origin, structure, and evolution of extremely dusty inner planetary systems.
\\*[1.0mm]}

\end{abstract}




\large{

\section{\large \bf Introduction}

Infrared observations of main sequence stars have demonstrated
the existence of exceptionally dusty inner planetary systems.
This is defined here to mean those stars hosting infrared excess emission having
fractional infrared luminosity ($\tau$$=$L$_{\rm IR}$/L$_{\rm bol}$)
of $\gtrsim$1\% for dust populations characterized by
blackbody emission with an effective temperature of T$_{\rm dust}$$\gtrsim$300\,K.
These are the dustiest main sequence stars known and such systems are exceedingly rare
(e.g., \citealt{uzpen09}; \citealt{balog09}; \citealt{melis10}; \citealt{kennedy13}).
To date, only a handful are known that have such high levels of 
mid-infrared excess emission and hence inner planetary system dust (see 
e.g., \citealt{gorlova04}; \citealt{song05}; \citealt{gorlova07}; 
\citealt{rhee07b}; \citealt{rhee08}; \citealt{melis10}; 
\citealt{zuckerman12}; \citealt{melis12}; \citealt{schneider13}; \citealt{melis13};
 \citealt{gaidos19}; \citealt{tajiri20}; \citealt{moor21}).
Although it seems reasonably settled that transient collisional events between 
rocky bodies are necessary to generate the exceptionally dusty disks
(\citealt{wyatt08}; \citealt{fujiwara12}; \citealt{olofsson12}),
it is not clear if the dust is generated in a specific 
star- or planet-formation event or how it might impact fully-formed planets 
(\citealt{melis16}; \citealt{kral17}; \citealt{moor21}). 

Some extremely dusty main sequence systems have recently 
been found to additionally exhibit dimming events due to the passage of circumstellar 
material along our line of sight (e.g., \citealt{dewit13}; 
\citealt{kennedy17}; \citealt{gaidos19}; \citealt{tajiri20}).
These are similar in lightcurve behavior to the ``dipper'' behavior seen for protoplanetary
disk systems (e.g., \citealt{moralescalderon11}; \citealt{cody14}, and references therein), 
although their circumstellar material is thought to be secondary in nature
(generated by the collisional breakdown of mature planetesimals/planets).
Such systems are typically young ($\lesssim$100\,Myr) and in at least one case are also host to
gaseous material (e.g., \citealt{punzi18}).

Main sequence systems with both strong infrared excess emission from inner planetary
system dust and ``dipper'' behavior can provide unique insight into the structure
and evolution of this material (e.g., \citealt{kennedy17}; \citealt{gaidos19}).
Infrared excess emission provides the means to localize (to some extent)
the dust in the planetary system (avoiding potentially confounding situations 
like in the cases of J140747.93$-$394542.6,
KIC\,8462852, or similar stars;
\citealt{mamajek12};
\citealt{boyajian16,boyajian18}; \citealt{meng17}; \citealt{david17}; 
\citealt{mentel18};
\citealt{ansdell19}; \citealt{rappaport19}; \citealt{saito19})
while occultations in the lightcurve
provide detailed information on its opacity and spatial organization (e.g., 
\citealt{vanwerkhoven14}; \citealt{kenworthy15};
\citealt{kennedy17}).

In this paper we present the discovery of the oldest ($\sim$600\,Myr) extremely dusty main 
sequence star to also host dimming events due to orbiting material.

\section{\large \bf Literature Summary}

TYC\,8830~410~1 was first discovered to be an infrared excess star in the survey
of \citet{cotten16}. They suggest an uncertain spectral type of K3 (stellar T$_{\rm eff}$ of 4900\,K),
blackbody-fit dust temperature of 425\,K
and associated orbital radius of 0.2\,AU for blackbody-emitting grains,
and a fractional infrared luminosity of 1.2\%.
Subsequent works also found TYC\,8830~410~1 to be an excess star
(e.g., \citealt{marton16}; \citealt{mcdonald17}), but no further characterization of the excess
was presented.

Optical spectroscopy of TYC\,8830~410~1 was conducted as part of the RAVE survey
\citep{kunder17}. From observations made on 2009-11-10
(MJD of 55145.39994213) they measured a heliocentric radial velocity of 
7.0$\pm$1.8\,km\,s$^{-1}$, stellar T$_{\rm eff}$ of 5350$\pm$140\,K,
stellar gravity log$g$ of 4.4$\pm$0.3 in cgs, and metallicity [M/H] of $-$0.1$\pm$0.2.
Further analysis on the RAVE spectra by \citet{zerjal17} suggested the presence of chromospheric 
Ca~II infrared triplet emission from which an age of $\approx$370\,Myr is estimated.

TYC\,8830~410~1 has appeared in every {\it Gaia} release. We adopt
parameters measured for it from DR2 and EDR3
\citep{gaia18,gaia20};
these are displayed in Table \ref{tabstarpars}.
While investigating the {\it Gaia} data, we identified a co-moving wide-separation
companion to TYC\,8830~410~1. We describe this object in Section \ref{seccomp}.

\section{\large \bf Observations}
\label{secobs}

In this section we describe observations obtained for this work and archival
data analyzed for this system for the first time.

\subsection{\large \bf FEROS}

Multiple epochs of optical echelle spectroscopy were obtained for TYC\,8830~410~1
with FEROS at the MPG/ESO 2.2\,m telescope at La Silla Observatory
\citep{kaufer99}.
Observations were conducted in the ``Object-Calibration'' mode with one fiber
obtaining a simultaneous ThAr lamp spectrum to produce precise ($\lesssim$20\,m\,s$^{-1}$)
radial velocities to aid in searching for close-separation companions.

Data are reduced with CERES \citep{brahm17}, which also produces precision radial velocities
for each epoch and associated uncertainties. FEROS observation epochs and measured velocities
are listed in Table \ref{tabrvs} along with literature velocities.

\subsection{\large \bf MagE}

Observations with MagE at Magellan/Baade were obtained for the wide-separation companion 
on UT 10 November 2019. The spectrograph was used with a 0.5$'$$'$ slit resulting in
R$\sim$8,000 spectra from 4100-10600 \AA . 
Data are reduced with the facility Carnegie Python pipeline \citep{kelson00,kelson03}.
A total integration time of 2700~seconds
resulted in a signal-to-noise ratio of 40 per pixel near H$\alpha$, 50 per pixel near Li~I $\lambda$6708, 
and $>$50 per pixel in the TiO bands at 7000-7600 \AA . 
Two RV and spectral type standard stars were also observed with the same setup: 
GJ\,54.1 in a 60~second integration and GJ\,908 in a 10~second integration.

\subsection{\large \bf VISIR}
\label{secvisir}

Mid-infrared imaging and spectroscopy were obtained with the VISIR instrument
(\citealt{lagage04}; \citealt{kaufl15})
mounted on VLT-Melipal at Paranal Observatory. Observations were conducted in service mode. 

Imaging observations were conducted on UT 04 January 2016 
in the AutoChopNod mode with default parameters, the chop/nod direction set to perpendicular, 
and positioning of the source in the left half of the chip. 
The 1024 $\times$ 1024 pixel detector was configured for 
0.045$''$\,pixel$^{-1}$ yielding a field of view of roughly 46$''$ $\times$ 46$''$.
Observations were performed with the PAH1 filter (8.59\,$\mu$m central wavelength and a 
half-band width of 0.42\,$\mu$m) and exposed for a total of 2440~seconds on source. 
The flux standard HD\,220440 \citep{cohen99} was observed
immediately after observations of TYC\,8830~410~1.

Spectroscopic observations were conducted over four nights in late September/early 
October 2016. Each visit typically consisted of observations of TYC\,8830~410~1
and two calibration stars (one before and one after observation of the science target), 
spending $\approx$90~minutes total wall clock time on
the science target. The detector was configured for low-resolution spectroscopy and
yielded a 0.076$''$\,pixel$^{-1}$ spatial scale and spectral resolving power of R$\sim$300 
from 7.5-14\,$\mu$m with the N-band prism and 0.75$'$$'$ slit.
Sources were nodded along the slit with a 10$''$ throw to cancel out
background emission and structure.

Data reduction was performed with in-house IDL routines optimized for background-limited
observations. In brief, two-dimensional
images were chop- and nod-differenced and combined for both the standard stars and 
the science target. 
Flux was extracted for each positive and negative beam for all targets with an
aperture that yielded approximately 85\% encircled energy. 

For imaging data, each of the four chop/nod beams were
averaged and the uncertainty set to the standard deviation 
of these four measurements divided by 2 (the
square root of the number of measurements). The VISIR calibration webpage-given 
flux density for HD\,220440 (9.78\,Jy)
seemed to be too low compared to satellite measurements, 
so we constructed a broad-band spectral energy distribution for the
star from available high-fidelity photometry and fit a stellar atmospheric model to it
(e.g., \citealt{cotten16}). From the model fit we
estimated the flux in the PAH1 filter band-pass to be 10.37\,Jy
and used that when flux-calibrating the extracted counts for TYC\,8830~410~1.
Photometric measurements for TYC\,8830~410~1, including from VISIR, are given
in Table \ref{tabphot}.

Spectroscopic data reduction followed that done for Subaru/COMICS data
in \citet{su20}. Slight changes were made to the code to account for different
spectrum projections onto the VISIR detector and the presence of two negative
spectral beams in the VISIR data;
only spectral samples with signal-to-noise ratio per pixel of $\gtrsim$3 were
kept for the final spectrum.
From the final combined mid-infrared spectral
data set we measure a signal-to-noise ratio of $\approx$10 near 10.5\,$\mu$m

\subsection{\large \bf {\it WISE} }

{\it Wide-field Infrared Survey Explorer} ({\it WISE}, \citealt{wright10}; \citealt{mainzer11};
\citealt{mainzer14}) epoch data products
are used to explore variability in the TYC\,8830~410~1 system in the thermal- to mid-infrared
(W1/3.35\,$\mu$m, W2/4.60\,$\mu$m, W3/11.56\,$\mu$m,
and W4/22.09\,$\mu$m channels).
The W3 and W4 channels only collected data over a short time period between
MJD 55,324-55,326.
Data were accessed via IRSA\footnote{\url{https://irsa.ipac.caltech.edu/Missions/wise.html}}
and are taken as reported.

\subsection{\large \bf ASAS-SN}

All-Sky Automated Survey for Supernovae (ASAS-SN) photometry data products 
(e.g., \citealt{shappee14}; \citealt{jayasinghe19})
were utilized
in assessing the history of dimming events toward TYC\,8830~410~1.
We additionally downloaded data products for stars nearby in the plane of the sky
to TYC\,8830~410~1 with comparable magnitudes for comparison purposes.

\subsection{\large \bf {\it TESS} }

{\it Transiting Exoplanet Survey Satellite} ({\it TESS}; \citealt{ricker14}) data are
available for TYC\,8830~410~1 in two sectors. Full-frame image (FFI) data
were obtained during observations of Sector~1 while 2~minute cadence
data were obtained during Sector~28.
A lightcurve from the FFI data products is obtained from the reductions performed
by \citet{huang20a,huang20b}. We additionally reject some data points based on quality flags,
unusual character within the time range of BJD 2,458,347-2,458,350, and an additional
buffer region on either side of the downlink gap (BJD 2,458,337.8-2,458,340.4).
Shorter cadence data are simple aperture photometry (SAP) flux values
as retrieved from MAST\footnote{\url{https://mast.stsci.edu/portal/Mashup/Clients/Mast/Portal.html}}.

\subsection{\large \bf LCOGT}
\label{seclcogt}

Las Cumbres Observatory Global Telescope (LCOGT; \citealt{brown13})
monitoring of TYC\,8830 410 1 has been ongoing
since April 2020. In this paper we present single-band monitoring data obtained through
the end of September 2020; multi-band monitoring data has been
obtained since then and will be presented in future works.

Images are acquired with the 0.4m robotic telescope network
and were requested to be obtained with a cadence of 2~hours. The 0.4m network
is a system of Meade 16-inch telescopes equipped with 
SBIG STL6303 cameras. The cameras host detectors with a plate scale of
0.571$'$$'$\,pixel$^{-1}$ and a field-of-view of 29$'$$\times$19$'$.
Throughout the 2020 observing period presented herein,
only two southern-hemisphere sites were operating $-$
Siding Spring Observatory in Australia and Sutherland Observatory in South Africa.
For each visit, telescope guiding was active and three images of 10~seconds each were
obtained of TYC\,8830~410~1 with the telescope pointing center within 30$'$$'$ of
the target star. Images were obtained in a Bessel V-band filter.

Data are reduced by LCOGT with the 
{\sf BANZAI} data pipeline (\citealt{mccully18}).
In brief, this pipeline corrects for bad pixels, subtracts bias and dark current,
performs flat-field correction, conducts source extraction with the 
{\sf SEP} software suite\footnote{\url{https://github.com/kbarbary/sep}},
and then attempts to obtain an astrometric solution with the methods
of \url{http://astrometry.net/}. Only images where a successful astrometric
solution is obtained are used in subsequent analysis.
Due to strongly variable seeing and telescope focus between epochs, we adopt
Kron-aperture \citep{kron80} magnitudes as produced by {\sf SEP} for the target and
comparison stars; this choice effectively ensures that the target and comparison stars
have apertures with similar encircled energy in every epoch.

A selection of stars within the LCOGT imaging field-of-view 
with similar brightness as TYC\,8830~410~1 
are used as comparison stars to derive magnitude measurements for the target star
in each epoch. Two comparison stars reproduce well each others' known V-band
magnitudes and show no obvious trends throughout the LCOGT monitoring period.
These stars have J2000 positions
of 23~00~11.84 $-$58~54~34.6 and 23~00~27.61 $-$58~57~32.4 and have
V-band magnitudes of 12.25 and 12.37, respectively (these are a combination of
ASAS-SN and {\it Gaia} results for both stars and have uncertainty of $\sim$1\%).
For each visit, we adopt the median value of the three measured magnitudes for
TYC\,8830~410~1 as the epoch measurement and the standard deviation
as the uncertainty. All measurements are reported in the Appendix in Table \ref{tabappA}.





\section{\large \bf Results}
\label{secmodres}

\subsection{\large \bf Stellar properties}
\label{seccomp}

As the FEROS spectra for TYC\,8830~410~1 do not display significant variability
(see below), we combine them all into one super-spectrum to characterize the star. 
From the line ratios of \citet{strassmeier90b} and \citet{padgett96} 
we determine that the star has an effective temperature of 5300$\pm$400\,K
and a spectral type of G9$\pm$2, respectively. These are consistent with the results
from analysis of RAVE spectra as discussed above, although less precise in general,
thus we adopt the RAVE spectroscopic-derived stellar effective temperature
and overall spectral type for the star of G9\,V (Table \ref{tabstarpars}).
In the FEROS super-spectrum we are unable to detect Li~I $\lambda$6708
absorption and set a 3$\sigma$ equivalent width limit of $<$7\,m\AA\ (Figure \ref{figosp}).
The bluest orders of the FEROS spectra are then searched for evidence of
chromospheric emission in the Ca~II H+K lines; no obvious core-reversal emission
is seen (Figure \ref{figosp}). 
Following the  methodology of \citet{hempelmann16} and references therein, 
we calculate an S-index of 0.26$\pm$0.01 from the FEROS super-spectrum
and convert it to log\,$R'$$_{\rm \it HK}$ of $-$4.75$\pm$0.10 following the description
in \citet{noyes84}.

{\it Gaia} proper motion and parallax measurements for stars in the field around
TYC\,8830~410~1 reveal a 49.5$'$$'$ separation co-moving and co-distant
companion. This star, 2MASS\,J23011901$-$5858262 (hereafter 2MASS\,J2301$-$5858),
has an absolute magnitude and G$_{\rm BP}$$-$G$_{\rm RP}$ color
(Table \ref{tabstarpars}) that strongly suggest it is a mid-M dwarf star.
MagE spectra of the companion (Figure \ref{figcomposp})
demonstrate it is consistent with a spectral type of M4\,V, 
has moderately strong Balmer H$\alpha$ emission (equivalent width of 2.1$\pm$0.1\,\AA ,
presumably due to magnetic activity), and has no detectable Li~I $\lambda$6708 absorption
(with a 3$\sigma$ equivalent width limit of $<$60\,m\AA ).
We additionally measure a radial velocity for the companion consistent with the
average of FEROS velocity measurements for the primary star (Tables
\ref{tabstarpars} and \ref{tabrvs}), thus further confirming them to be truly bound.

Radial velocity data from FEROS (and to a lesser degree other available radial velocity
measurements in Table \ref{tabrvs}) are used to search for and constrain the presence
of other companions to TYC\,8830~410~1.
Taken at their quoted uncertainties, the FEROS velocities suggest variability is present
at the level of $\sim$0.1\,km\,s$^{-1}$. A periodogram search of the FEROS
velocities does not indicate any significant signals are present, but are suggestive of
a possible period around $\approx$4~days. 
Continued radial velocity monitoring of TYC\,8830~410~1 can determine if this signal is real
and whether it is due to stellar activity, possibly a hot Jupiter-like companion,
or maybe due to transiting dust clouds as described in \citet{dodin21}.
With no significant signals present in the FEROS velocity data, we conclude that
there are no stellar-mass nor massive sub-stellar
companions within an AU of TYC\,8830~410~1.


\subsection{\large \bf  Infrared excess emission}
\label{secdusres}

We revisit the infrared excess parameters presented in
\citet{cotten16} based on updated stellar and infrared measurements presented herein.
Figure \ref{fig8830sed} shows that this system hosts warm inner planetary system 
dust with strong solid-state emission resulting in a 
fractional infrared luminosity of $\sim$1\%, easily placing
it in league with other exceptionally dusty main sequence star systems (e.g., 
\citealt{melis16}; \citealt{moor21}). 
High spatial resolution VLT/VISIR mid-infrared imaging observations find only
a single point source, indicating that
this object is the source of the {\it WISE}-detected excess flux, thus confirming it
as a bona-fide exceptionally dusty inner planetary disk system.

Stellar parameters retrieved from the spectral energy distribution fit suggest
an effective temperature of $\approx$5,000\,K which is lower
(albeit not especially significantly) than
the spectroscopically-retrieved value of 5,350\,K. As discussed below, the star
is likely to be seen through a varying level of dust and thus could be reddened
leading to the lower spectral energy distribution fit-value.
We attempted a second stellar spectral energy distribution fit with the model effective
temperature fixed at 5,300\,K and found a reddening of A$_{\rm V}$$=$0.5\,mag
with the \citet{cardelli89} extinction curve (and R$_{\rm V}$$=$3.1) provides
a reasonable fit to the stellar photometry. Additional evidence for reddening
appears in the color-magnitude diagram position of TYC\,8830~410~1
as discussed in Section \ref{secage}.

The VISIR mid-infrared spectrum (Figures \ref{fig8830sed} and \ref{figvisir})
clearly reveals a strong solid-state emission
feature with the characteristic amorphous and crystalline silicate peaks
at $\approx$10 and 11\,$\mu$m (e.g., \citealt{honda04}, \citealt{chen06}, \citealt{lisse08}).
Fitting proceeds as in \citet{weinberger11} and \citet{olofsson12}, which we briefly
summarize here. We used the same optical constants and absorption coefficients as in
\citet{olofsson12}; these are taken 
from \citet{Dorschner1995} and \citet{Jaeger2003} for amorphous silicates with olivine and pyroxene stoichiometry, 
from \citet{Tamanai2010} for the ``Mg-rich'' and ``Fe-rich'' crystalline olivine, 
from \citet{Tamanai2010a} for the silica, 
from \citet{Jaeger1998} for the crystalline enstatite, and 
from \citet{Jaeger1998b} for the carbonaceous dust grains. 
For all the dust species, the minimum allowed grain size is $0.1$\,$\mu$m while the 
maximum grain size is set to $1$\,$\mu$m for the crystalline dust grains and $1$\,mm 
for the amorphous dust species. 
The free parameters of the modeling are the inner radius and radial width of the dust ring, 
the slope of the density distribution, and the slope of the grain size distribution. 
The radial distribution is sampled over $n_\mathrm{r} = 80$ bins, and for each dust species 
and each grain size the temperature of the particles at a given distance is calculated by 
equating the energy received and emitted (Eqn.\,3 of \citealp{olofsson12}). 
For each dust species, we then compute an emission profile at the same wavelengths 
as the observations, weighted by the grain size distribution. 
For a given set of free parameters, we then find the linear combination of the emission 
profiles that best reproduces the observed spectrum, using the 
\texttt{lmfit} package (\citealp{lmfit}). To find the best solution, we used 
the \texttt{Multinest} nested sampling algorithm (\citealp{Feroz2009,Feroz2019}), 
interfaced with \texttt{Python} using the \texttt{PyMultiNest} package (\citealp{Buchner2014}).

From the available spectrum we are able to constrain 
the inner radius of the disk of emitting dust grains to be $\sim$0.25\,AU (and certainly $<$0.5\,AU)
and the crystallinity fraction to being $\sim$30\% by mass. We are not able to robustly
constrain the type of emitting grains (although some combination of forsterite and enstatite
most likely contribute to the 11\,$\mu$m shoulder) 
and that the grain size distribution favors small ($<$2\,$\mu$m) dust particles.
A representative fit illustrating these dust properties is shown in Figure \ref{figvisir}.

It is found that the dust continuum temperature is not well-constrained
as most available excess measurements are part of the solid state emission complex
(we assume there are strong silicate emission features near $\sim$20\,$\mu$m that enhance
the {\it WISE} $W4$ measurement). In arriving at
the continuum curves plotted in Figure \ref{fig8830sed}, we require the fit to pass through
the bottom of the high- and low-wavelength ends of the VISIR N-band spectrum.
In doing so, we arrive at temperatures T$_{\rm dust}$ of 300-350\,K.

\subsection{\large \bf Occultations}
\label{secoccult}

TYC\,8830~410~1 was observed with {\it TESS} in Sector 1 as a
Full-Frame Image target (30~minute cadence).
The beginning of the {\it TESS} lightcurve just 
catches a $\sim$1~magnitude deep, $\approx$1.5~day duration eclipse
with irregular shape (Figure \ref{figtess}).
Also evident is general variability with a stochastic
nature to it (no identifiable period within the $\approx$28-day timespan).
Examining precovery ASAS-SN lightcurve data for TYC\,8830~410~1 shows
several dips compatible with the depth of the {\it TESS}-observed feature
as well as stochastic variability that results in a lightcurve rms deviation from
the mean of 
$\approx$0.14\,magnitudes (Figure \ref{figasassnlco}). 
A periodogram for the ASAS-SN data does not
reveal significant signals at any period. Phase-folding the lightcurve across
a range of values reveals that some periods between 120-200~days 
can line up most of the dips, but not all of them.

ASAS-SN lightcurve data for stars nearby and of comparable brightness to 
TYC\,8830 410 1 do not show any dips nor the stochastic variability 
(ASAS-SN check stars showed rms fluctuations in their lightcurves at the 
0.01-0.02~magnitude level). We also examined {\it TESS} data for similar magnitude 
stars around TYC\,8830~410~1 and did not find any others with comparable features.
This leads us to conclude that the features observed in the optical data 
are astrophysical and not instrumental in nature.
{\it WISE} epoch photometry spanning precovery to post-{\it TESS} Sector 1 epochs
are suggestive of possible deep eclipses even in the thermal infrared.
However, after comparison with nearby stars of similar magnitude as
TYC\,8830~410~1, it is determined that apparent dips are actually
instrumental in nature.

Based on the {\it TESS} eclipse feature and support for occultations from ASAS-SN,
we pursued additional {\it TESS} data in the extended mission with a higher
cadence and ground-based routine monitoring with LCOGT.
LCOGT monitoring shows a wide range of variability, including $\sim$1~day duration
deep eclipses and a host of smaller depth features of various durations
(Figure \ref{figasassnlco}). 
The LCOGT data display an rms deviation from the mean of $\approx$0.17\,magnitudes,
reasonably consistent with that seen in the ASAS-SN data (meaning no
obvious evolution of the dust screen has occurred in the $\sim$6.5 years
of observations presented herein).
The 2~minute cadence lightcurve from {\it TESS} Sector 28 
overlaps with LCOGT monitoring and similarly shows highly structured 
variability (Figures \ref{figtess}).
In general the appearance of {\it TESS} and 
LCOGT/ASAS-SN data where they overlap are similar, although
there are some disagreements (especially in the absolute flux level)
that could be instrumental or possibly astronomical in nature ({\it TESS} has
a redder bandpass than the V-band monitoring done with LCOGT and ASAS-SN).
Ongoing and future ground-based multiband observations can reveal if the depth
of eclipse features are wavelength dependent.

We group the types of dimming events for TYC\,8830~410~1 into three categories.
In the first category is the stochastic variability when the star is between V-band
magnitudes of 11.4 to 12.1. The brightest V-band magnitude measured with ASAS-SN or LCOGT
for TYC\,8830~410~1 is 11.46$\pm$0.01, which we take to be the unocculted V-band
magnitude.
In the second category are medium dimming events
where the star fades to V-band magnitudes of 12.2 to 12.3. The third category
is for the deep dimming events when the star is extinguished to V-band magnitudes
of fainter than 12.3. Each of ASAS-SN, {\it TESS} (in the two different sectors),
and LCOGT see all three categories in their monitoring data.

\section{\large \bf Discussion}
\label{secdisc}

\subsection{\large \bf System age}
\label{secage}

Estimates for the age of the TYC\,8830~410~1 system must be self-consistent for 
both stars. In addition to measurements described above, we also include an X-ray upper limit
from {\it ROSAT} \citep{truemper82} and {\it eROSITA} 
(extended ROentgen Survey with an Imaging 
Telescope Array; \citealt{predehl21})
for TYC\,8830~410~1 (and technically its companion, although that
is less constraining), 3D Galactic space kinematics (Table \ref{tabstarpars}), and
simultaneous isochrone fits for both stars.

Lithium limits for TYC\,8830~410~1 rule out ages $\leq$100\,Myr and suggest
an age $\gtrsim$200\,Myr (e.g., \citealt{zuckerman04}). 
The non-detection of lithium in the companion spectrum is consistent
with these age bounds, although in and of itself is not especially constraining
(suggesting ages $\gtrsim$20\,Myr).

Analysis of RAVE Ca~II infrared triplet data suggested
an age of $\approx$370\,Myr \citep{zerjal17}. Our analysis of Ca~II H+K
activity levels from the FEROS data suggest an age range of 1.5-4.0\,Gyr if one applies
directly Equation 3 of \citet{mamajek08}. However, if one instead considers
Figures 4 and 5 and Tables 7 and 9 of \citet{mamajek08}
it is possible $-$ given the range of log\,$R'$$_{\rm \it HK}$
values observed for various open cluster members of known age $-$
that TYC\,8830~410~1 could be between 100\,Myr and 4\,Gyr. The extremes of the latter
age range
would require TYC\,8830~410~1 to be an activity outlier while ages between
600\,Myr to 2\,Gyr are more likely to produce the observed activity level. As such, we consider
chromospheric activity to rule out ages $<$100\,Myr and to be suggestive of an 
age $\gtrsim$500\,Myr. Chromospheric activity for the companion star in the form of H$\alpha$ 
emission is consistent with chromospheric ages suggested for the
primary star, and specifically suggests an age $\lesssim$2\,Gyr (e.g., \citealt{shkolnik09};
\citealt{kiman21}, and references therein).

The {\it ROSAT} all-sky survey and first {\it eROSITA} all-sky survey
did not detect TYC\,8830~410~1
nor its companion in the X-rays (A. Merloni 2021, priv.\ comm.). 
{\it eROSITA} provides better sensitivity than does {\it ROSAT},
so we focus discussion on its X-ray limits.
We obtain X-ray limits by assessing the flux and associated uncertainties for detected
sources in the region of TYC\,8830~410~1. A conservative limit of the $\approx$95\%
source completeness level is adopted, resulting in
a flux limit in the 0.6-2.3\,keV band of $<$2$\times$10$^{-13}$\,ergs\,cm$^{-2}$\,s$^{-1}$. 
The {\it eROSITA} limit suggests a ratio of X-ray to bolometric luminosity for 
TYC\,8830~410~1 of log(L$_{X}$/L$_{\rm bol}$)$<$$-$3.6 and a limit for its companion of
log(L$_{X}$/L$_{\rm bol}$)$<$$-$2.0. Comparison to Figure 4 of \citet{zuckerman04}
suggests the primary is $\gtrsim$100\,Myr old.
Limits for the companion are not especially restrictive
(e.g., \citealt{zuckerman04}; \citealt{stelzer13}).

Three-dimensional Galactic space motions ({\it UVW} in Table \ref{tabstarpars})
are compatible with young stars ($\lesssim$100\,Myr) within 200\,pc of the Sun
(e.g., \citealt{zuckerman04,torres08}). Interestingly, the best matches between the
{\it UVW} space motions for the TYC\,8830~410~1 system and known nearby moving
groups are $\epsilon$\,Cha and the Local Association or Pleiades moving group. 
The former is far too young to be home
to TYC\,8830~410~1 (with an age of $\sim$6\,Myr), in addition to
inconsistencies between their respective distances and locations on the plane of the 
sky (\citealt{torres08}).
The legitimacy of the latter association is questionable (e.g., \citealt{zuckerman04}) and
in any case is suggested to mostly have ages between 20-150\,Myr
(\citealt{montes01} and references therein), again incompatible
with age constraints for TYC\,8830~410~1 discussed above.
Other possible matches also suffer distance, age, or positional issues
(e.g., \citealt{gagne18}; \citealt{baluev20}).
TYC\,8830~410~1 appears to have a young-star-like {\it UVW} space motion
by coincidence, but otherwise is not particularly young (as suggested by other age indicators).
In such a case, kinematics are not capable of placing any reasonable constraints on the age
of the system.

Attempts to age-date the TYC\,8830~410~1 system via simultaneous isochrone fitting
are confounded by what appears to be reddening of the primary star. In most color-absolute
magnitude diagrams TYC\,8830~410~1 appears either too bright or too red
(e.g., Figure \ref{figiso}).
Based on the evidence for significant (and variable) quantities
of dust lying along our line of sight to the primary star, we consider it to be reddened. 
We do, however, mention briefly and dismiss the possibility that TYC\,8830~410~1
itself could be an unresolved binary system composed of nearly equal-mass stars. This 
setup is highly contrived as it would require the binary orbit to not exhibit obvious radial
velocity variability (a nearly face-on orbital inclination or wide orbit;
see restrictions in Section \ref{seccomp})
and to have orbital parameters that allow the inner disk component to exist over decade
timescales (see Section \ref{secoccult}).

Given the issues with the primary discussed above, 
we rely on the position of the companion alone to inform isochrone age estimates. In all
colors explored (Figure \ref{figiso}), 
2MASS\,J2301$-$5858 lies above the locus of field stars of similar spectral
type and tends to agree well with colors and absolute magnitudes of mid-M-type Hyades
stars. As such, we adopt isochrone age bounds of $>$500\,Myr and $<$5\,Gyr.

Taking in aggregate all of the above age bounds, we can confidently rule out ages $<$200\,Myr
for the TYC\,8830~410~1 system. The best age estimate is roughly Hyades-aged ($\approx$600\,Myr),
but the system could very well be between 500\,Myr-2\,Gyr in age.
X-ray detections of both stars could help improve the age estimate for this system, as well as
high signal-to-noise ratio optical spectroscopy monitoring the Ca~II H+K chromospheric activity
(especially if a stellar rotation period could be measured).

Based on the best age estimate above, 
TYC\,8830~410~1 is intermediate in age relative to other known extremely dusty
main sequence stars which span from several Myr to $>$1\,Gyr
(e.g., \citealt{melis16}; \citealt{moor21}, and references therein).
Notably, most extreme debris disk systems have ages $\lesssim$200\,Myr with the
exceptions of BD+20 307 (age $>$1\,Gyr, \citealt{zuckerman08b}; \citealt{moor21})
and TYC\,4479 3 1 (age 5$\pm$2\,Gyr, \citealt{moor21}). The latter system we consider
to be contaminated by Galactic dust emission (see Figure 1 of \citealt{moor21}
where extended nebular emission overlapping with the star is clearly evident
in the {\it WISE} channels where excess is claimed); it needs to be confirmed with 
higher resolution mid-infrared imaging before being included in any extremely
dusty main sequence star analyses.
As such, TYC\,8830~410~1 and BD+20 307 are the oldest (confirmed) extremely dusty
main sequence stars known and TYC\,8830~410~1 the oldest such system where
the dusty debris eclipses the host star (robust ages are not yet known for systems
presented by \citealt{tajiri20}).

TYC\,8830~410~1 joins the growing number of extremely dusty main sequence stars
with wide-separation binary companions. This association was first noted by
\citet{zuckerman15} and expanded upon in \citet{moor21}. The assertion by \citet{moor21}
that such dusty systems are more likely to host wide separation companions as a function
of age is supported by the age estimate for TYC\,8830~410~1 derived here.
\citet{moor21} develop a cometary delivery model to explain the origin
of extremely dusty main sequence stars $-$ especially the older population $-$
and suggest perhaps such delivery is amplified by instabilities
due to wide separation companions. 
The cometary model proposed by \citet{moor21} suffers from a major weakness
in that any instability which sends a significant quantity of mass
to a star's inner planetary system should also produce a substantial population of small dust
grains in the star's outer planetary system (e.g., \citealt{fujiwara12b}; \citealt{bonsor13,bonsor14};
\citealt{raymond14}, and references therein). Indeed, \citet{moor21} acknowledge this
shortcoming of their proposed cometary model and note results by \citet{vican16} which show very
few extremely dusty main sequence stars host outer planetary system dust populations
consistent with a cometary model.
In Section \ref{seccon} we suggest an alternative model that will be explored further in later works.

\clearpage
\subsection{\large \bf Dust properties}

With a fractional infrared luminosity of $\sim$1\%, TYC\,8830~410~1 is not especially
remarkable amongst the currently known sample of extremely dusty main sequence stars.
The dust composition is at first glance reasonably compatible with 
what is seen for other such stars (e.g., \citealt{olofsson12} and references therein),
although the available data leaves ambiguous many of the dust properties.
One potentially interesting difference that remains to be conclusively measured
is an apparent enstatite dominance over forsterite. In some of our models there can
be as much as a factor of 2 more enstatite than forsterite which is not typically
seen in extremely dusty star disks (e.g., \citealt{fujiwara10}; \citealt{olofsson12}). 
Enstatite-rich bodies are known in the solar system, including the surface of
Mercury and E-type asteroids which make up a significant fraction
of the inner asteroid belt
(e.g., \citealt{sprague98}; \citealt{zellner77}; \citealt{keil89}).
Future comprehensive mid-infrared spectroscopy (e.g., with {\it JWST}) can help
establish if this result is robust for TYC\,8830~410~1.

With the photometric monitoring conducted to date it is not possible to
demonstrate if the deeper dips seen are (quasi-)periodic or aperiodic.
Routine monitoring at a cadence of $\lesssim$1 day without gaps in temporal coverage 
is essential to catch these events or conclusively say they did not occur. For
example, the sporadic cadence of the ASAS-SN monitoring could
easily have missed any number of $\sim$1.5~day duration events like those
seen in the {\it TESS} and LCOGT data. Two such events are found in
the LCOGT monitoring from April to September 2020 separated by about 82~days,
although the April event was captured with a single epoch only and sufficient gaps 
in coverage are present that other events could have been missed.
Even allowing for missed events, it is not possible to find a periodic spacing
that matches the four lowest measured magnitudes in the
ASAS-SN data, the deep dip in the {\it TESS} data,
and the two deep dips in the LCOGT data. A period near 180~days can
get close to lining up most of the ASAS-SN and first LCOGT deep dips, but not the
{\it TESS} and second LCOGT deep dips which appear to themselves be separated
by a factor of roughly 180~days.
A speculative idea is that we might be seeing multiple orbiting sites of major 
collisions that shear out and disperse with time.

Stochastic variability similarly has no discernible period associated with it.
This combined with a lack of strong magnetic activity on TYC\,8830~410~1
indicates that this variability must also be due to dust transiting across
the face of the star. The medium dimming events are a few times more frequent than
deep dimming events and tend to be longer in duration
($\sim$3~days in the {\it TESS} Sector 28 lightcruve),
but again show no clear periodicity in the available data.

We conclude that all variability seen in optical lightcurves for TYC\,8830~410~1
is due to transiting dust. It is not possible to robustly identify the configuration of such
dust with the available data, but we comment on two possibilities.
In one configuration, the dust disk is vertically thin (scale height $\lesssim$ the stellar
diameter), radially narrow ($\Delta$R$_{\rm dust}$ $<$0.1\,R$_{\rm dust}$),
and the dust is fairly homogeneous in density in the vertical and radial axes.
In this case all the lightcurve changes are due to changes in the azimuthal 
density in the dust and indicate substantial clumpiness in the dust ring.
Such a configuration would be reminiscent of the distribution of material
presented in \citet{watt21} for post-giant impact-type events. Adapting the models
of \citet{watt21} to predict stellar brightness changes if post-collision dusty ejecta 
transits the host star would be valuable in further assessing the nature of TYC\,8830~410~1
and possibly the systems presented by \citet{gaidos19} and \citet{tajiri20}.

In a another $-$ perhaps more contrived $-$ 
configuration, the dust disk is again radially narrow and the density
is homogeneous in the radial and azimuthal axes. The disk has a structured
vertical density distribution and is additionally warped and precessing. Such a
configuration has been suggested for pre-main sequence ``dipper'' stars
where gas and/or interactions with stellar magnetic fields help shape the disk
inner edge (e.g., \citealt{bouvier03} and references therein). However, 
there is currently no evidence for gas in the disk around TYC\,8830~410~1,
and it appears to lie at orbital separations well beyond the reaches of the stellar
magnetic field (Section \ref{secdusres}). As such, any vertical structuring in its disk
(if present) would have to come from other sources, perhaps from Kozai-Lidov-type
interactions with the wide M-type companion. In this case the changes in the lightcurve
are due to the different heights in the disk being probed by the line of sight to the star. 
They would not be strictly periodic due to the precession of the warp and possibly due to
warp evolution.

\section{\large \bf Conclusions}
\label{seccon}

We present detailed characterization of the TYC\,8830~410~1 system. 
Infrared excess emission from circumstellar
dust is seen and occultations from this dust are caught in stellar photometric monitoring.
TYC\,8830~410~1 appears to have an age of $\sim$600\,Myr, and is definitely older
than typical extremely dusty main sequence stars which have ages of $\lesssim$200\,Myr.

The unusual deep dimming event shape, 
lack of a companion detection in radial velocity measurements,
the general stochastic variability seen in lightcurves, and the strong mid-infrared excess emission
with clear solid-state emission from small dust grains
point to a possible origin of the transit events as due to dust released in the aftermath of a
giant impact between rocky planets.
Giant impact-type collisions would produce significant quantities of
dusty ejecta that would go into orbit around the host star (e.g., \citealt{melis10};
\citealt{jackson12}; \citealt{genda12}; \citealt{watt21}, and references therein). 
Ejecta would collisionally grind itself
down generating small dust grains that would produce the observed mid-infrared excess
emission and the stochastic variability (similar to what is seen in \citealt{gaidos19}). 
Clumps of dust, or dust around the post-impact
rocky planet (possibly in a proto-lunar disk configuration; e.g., \citealt{kokubo00}) 
could be responsible for the deep dips seen if they subtend a 
large angular size relative to the host star
they orbit (e.g., \citealt{mamajek12}; \citealt{kenworthy15}).

The implied giant impact-type event is unlikely to be associated with
rocky planet formation
given the intermediate age of TYC\,8830~410~1 (such events should occur for stellar
ages $\lesssim$100\,Myr; e.g., \citealt{hartmann75}; \citealt{genda15}; \citealt{levison15}) 
and instead could possibly be 
due to a late-stage instability (e.g., \citealt{izidoro19}; \citealt{moor21}
and references therein).
\citet{izidoro19} specifically follow the long-term dynamical evolution of planetary systems
that form as chains of first order mean motion resonances (resonant chains),
finding that up to $\approx$95\% of resonant chains 
become dynamically unstable after dispersal of the gas disk.
They find timescales for these instabilities that extend up to the limit of their simulations
($\approx$300\,Myr), and it could be the case that such instabilities could occur for even
older ages.
Eccentricities and orbital inclinations of planets in a resonant 
chain grow in the absence of the damping effects of the gas disk
due to mutual interactions and their orbits can eventually cross
leading to collisions and scattering events. Eccentricities and
relative velocities could be further enhanced by the effects of a widely separated
companion (like 2MASS\,J2301$-$5858 in the case of the TYC\,8830~410~1 system)
through the Kozai-Lidov mechanism (e.g., \citealt{nesvold16} and references therein)
increasing the likelihood of instability and destructive collisions.
This instability-driven phase of late giant impacts
would be where the dust seen in extremely dusty stars like TYC\,8830~410~1
originates from.

If the proposed interpretation for TYC\,8830~410~1 is correct, then it would serve as a 
Rosetta stone for understanding exceptionally dusty stars. It would be capable of
providing the first-ever detailed
look at the structure of post-giant impact ejecta and its evolution, thus allowing direct
tests and constraints on models of this important pathway for rocky planet evolution.
Multi-band monitoring of this system is essential in conducting such work,
and with a reasonable brightness of V$_{\rm mag}$$\sim$12 amateurs could
easily contribute.


\acknowledgments

C.M.\ acknowledges support from NASA grants 18-ADAP18-0233
and 20-XRP20\_2-0038.
J.\ O. acknowledges support by ANID, -- Millennium Science Initiative Program -- NCN19\_171, from the Universidad de Valpara\'iso, and from Fondecyt (grant 1180395).
GMK is supported by the Royal Society as a Royal Society University Research Fellow.
M.K.\ acknowledges support by DFG grant KR 3338/4-1.
This research has made use of NASA's Astrophysics Data System, the SIMBAD database,
and the VizieR service.
This  publication  is  based  in  part  on  data  obtained  under 
CNTAC programs CLN2020A-007 and CLN2020B-002.
Based on observations collected at the European Southern Observatory under ESO programmes 296.C-5027(A) and 098.C-0427(A).
This paper includes data collected by the {\it TESS} mission, which are publicly available from the Mikulski Archive for Space Telescopes (MAST).
Funding for the {\it TESS} mission is provided by NASA's Science Mission directorate.



\facilities{VLT(VISIR), MPG(FEROS), Magellan(MagE), {\it TESS},  LCOGT}





}



\clearpage

\begin{figure}
 \centering
 \begin{minipage}{155mm}
  \includegraphics[width=155mm]{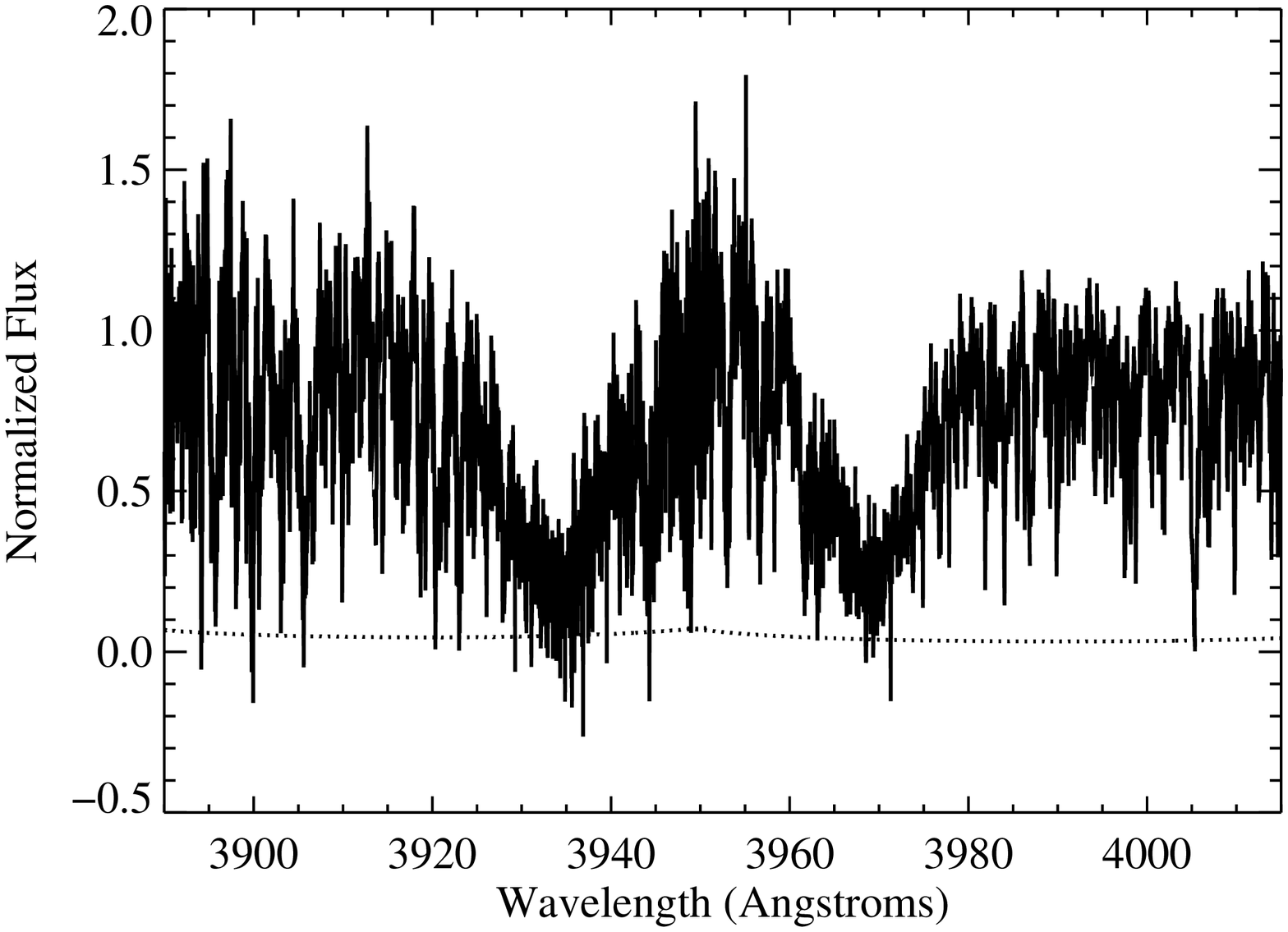}
 \end{minipage}
 \\*
 \begin{minipage}{155mm}
  \includegraphics[width=155mm]{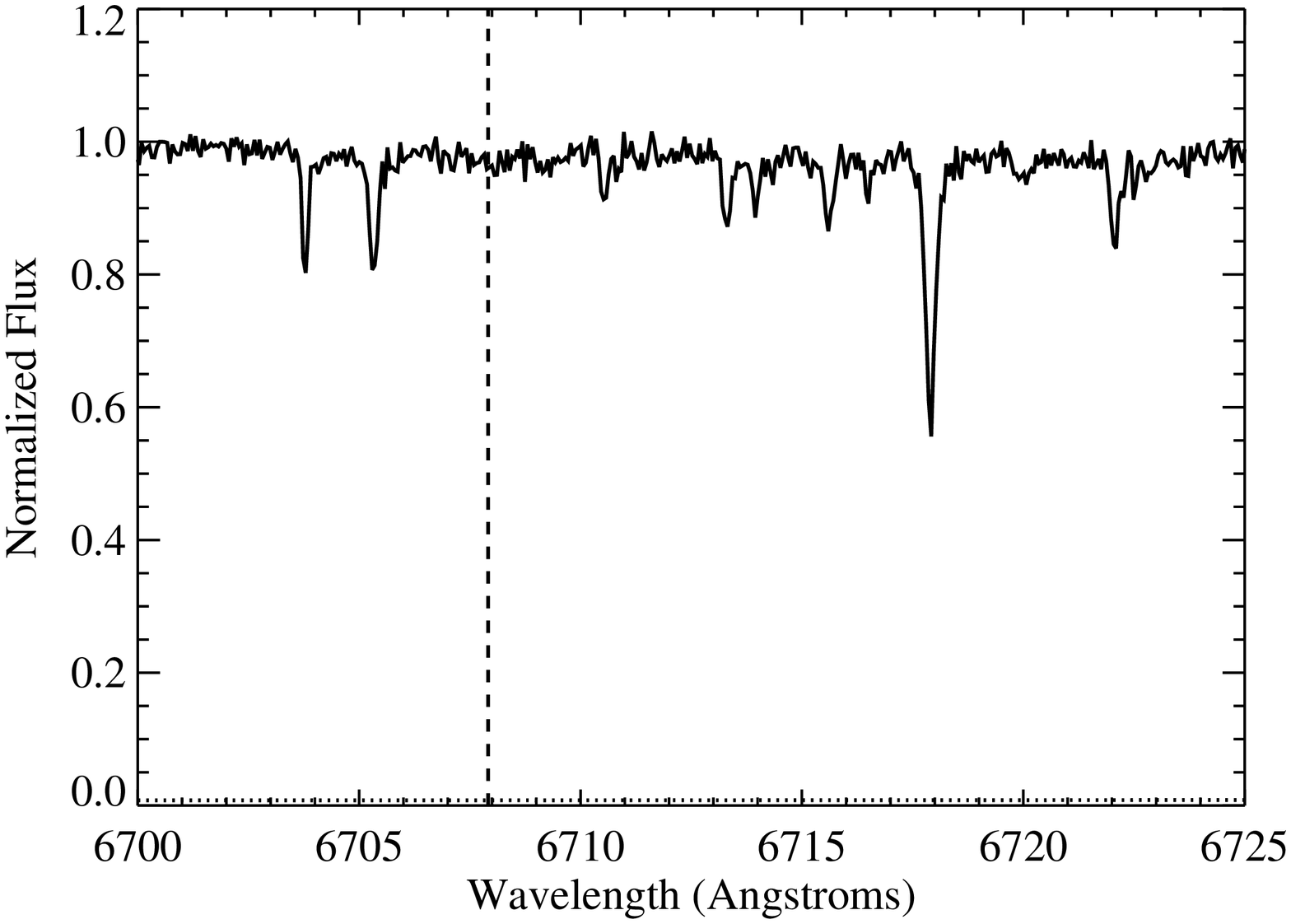}
 \end{minipage}
 \caption{\label{figosp} \large{FEROS super-spectra of 
              TYC\,8830~410~1 showing the Ca~II H+K ({\it top}) and Li~I $\lambda$6708
              ({\it bottom}) regions.
              Wavelengths are in air and corrected to the heliocentric reference frame,
              error spectra are plotted as dotted lines.
              The vertical dashed line in the bottom panel
              indicates the expected location of Li~I $\lambda$6708; no line is seen.} }
\end{figure}

\begin{figure}
 \centering
  \includegraphics[width=160mm]{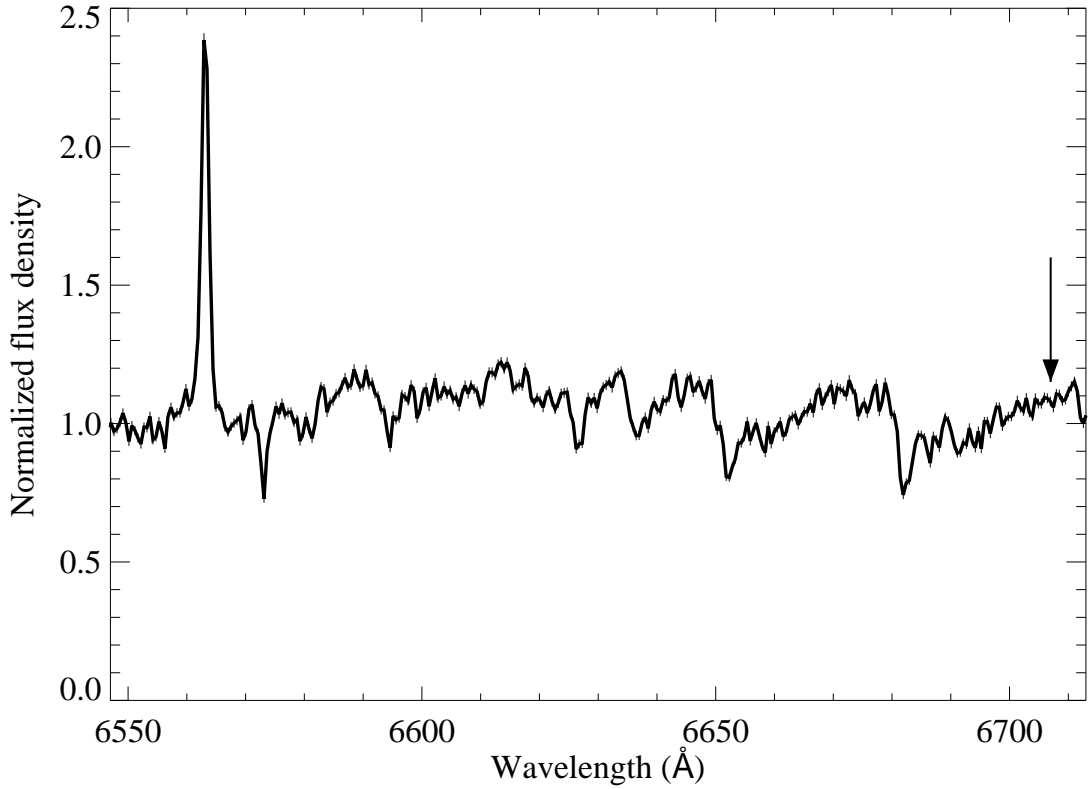}
 \caption{\label{figcomposp} \large{Segment of MagE spectra of 2MASS\,J2301$-$5858, the wide
              late-type companion to TYC\,8830~410~1. Normalized flux per pixel and associated
              error (thin vertical bars) are plotted. H$\alpha$ emission is evident, as well
              as molecular features typical of late-type stars. The downward-pointing arrow
              indicates the expected location of Li~I $\lambda$6708; no line is seen.
              Equivalent widths for H$\alpha$ and lithium are given in Section \ref{seccomp}
              while the companion spectral properties are given in Table \ref{tabstarpars}.} }
\end{figure}

\begin{figure}
 \centering
 \begin{minipage}{80mm}
  \includegraphics[width=83mm]{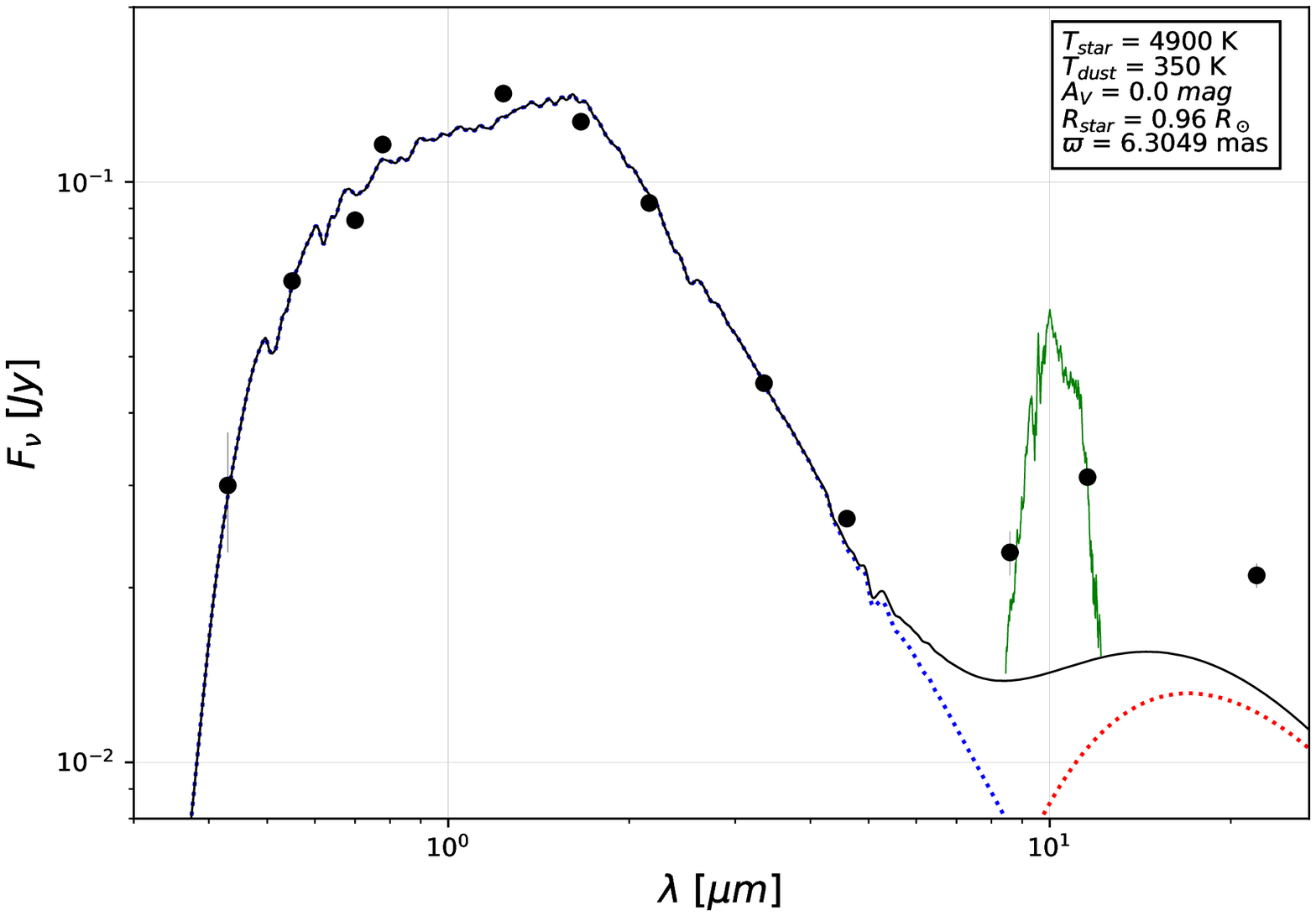}
 \end{minipage}
 \begin{minipage}{80mm}
  \includegraphics[width=83mm]{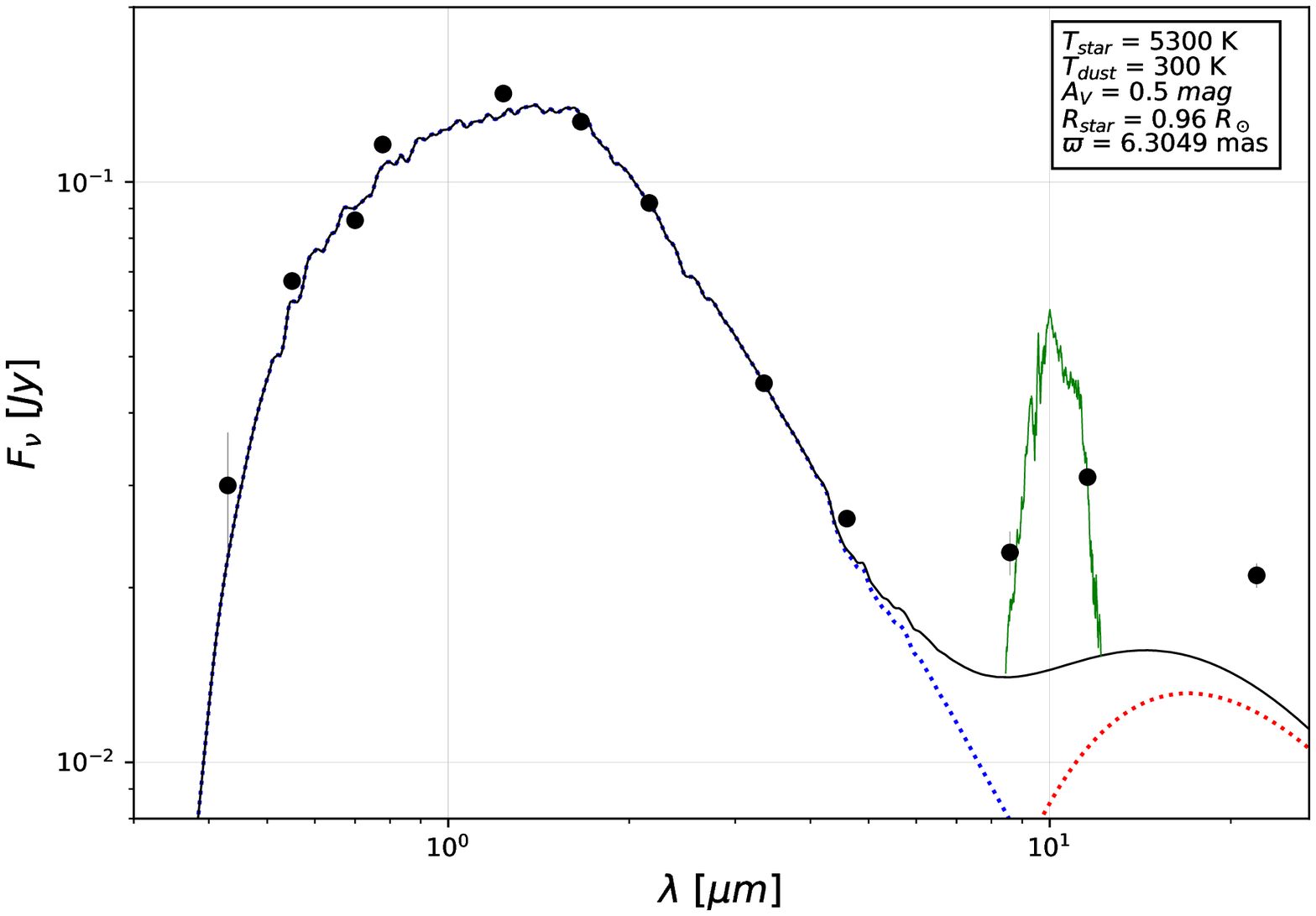}
 \end{minipage}
 \caption{\label{fig8830sed} \large{Spectral energy distributions for
                TYC\,8830~410~1 for two cases: no reddening ({\it left}) and
                reddening such that A$_V$$=$0.5 ({\it right});
                see discussion in Sections \ref{secdusres} and \ref{secage}. Data points are 
                as given in Table \ref{tabphot} and the green curve is the VISIR N-band spectrum.
                Vertical lines in data points indicate the measurement
                uncertainty. Some measurement uncertainties are smaller
                than the point sizes on the plot. The vertical scaling of the VISIR N-band spectrum
                is calibrated with the {\it WISE} $W3$ channel
                data point.
                The dotted blue curve connecting the $BVRIJHK_{\rm s}W1$ data points
                is a synthetic stellar atmospheric spectrum \citep{hau99}. 
                The dotted red curve is a blackbody at the temperature indicated
                on the figure panel; this temperature is not well-constrained
                as discussed in Section \ref{secdusres}. The solid black curve is the sum of
                the atmospheric and blackbody models.} }
\end{figure}

\begin{figure}
 \centering
 \includegraphics[width=150mm]{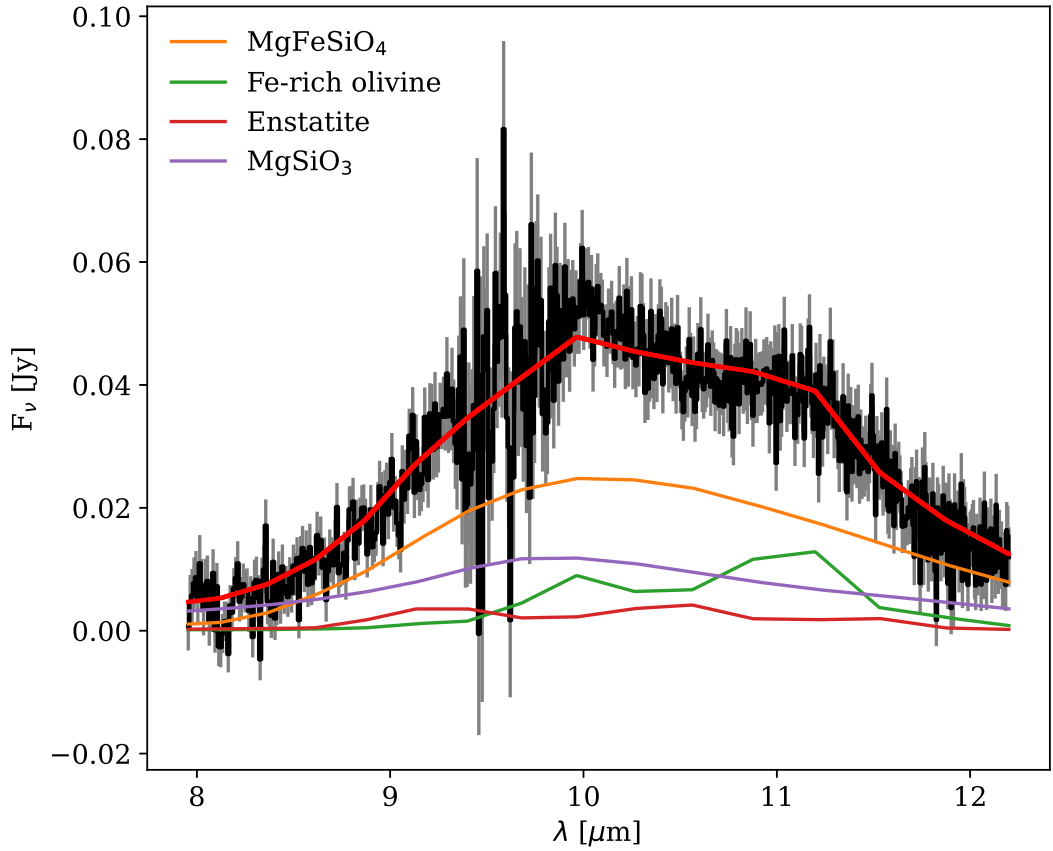}
 \caption{\label{figvisir} \large{VISIR N-band spectrum and representative model fit
                (see Section \ref{secdusres} for more details). The solid black curve is the measured 
                flux and the grey curve underneath is the associated uncertainty for each spectral
                sample. The red curve plotted over the data is the model fit which is the sum
                of the components plotted below the data and labeled in the figure legend.
                Fe-rich olivine (forsterite) and enstatite are crystalline species while MgFeSiO$_4$
                and MgSiO$_3$ are amorphous species.} }
\end{figure}

 \begin{figure}
 \centering
 \begin{minipage}{155mm}
  \includegraphics[width=155mm]{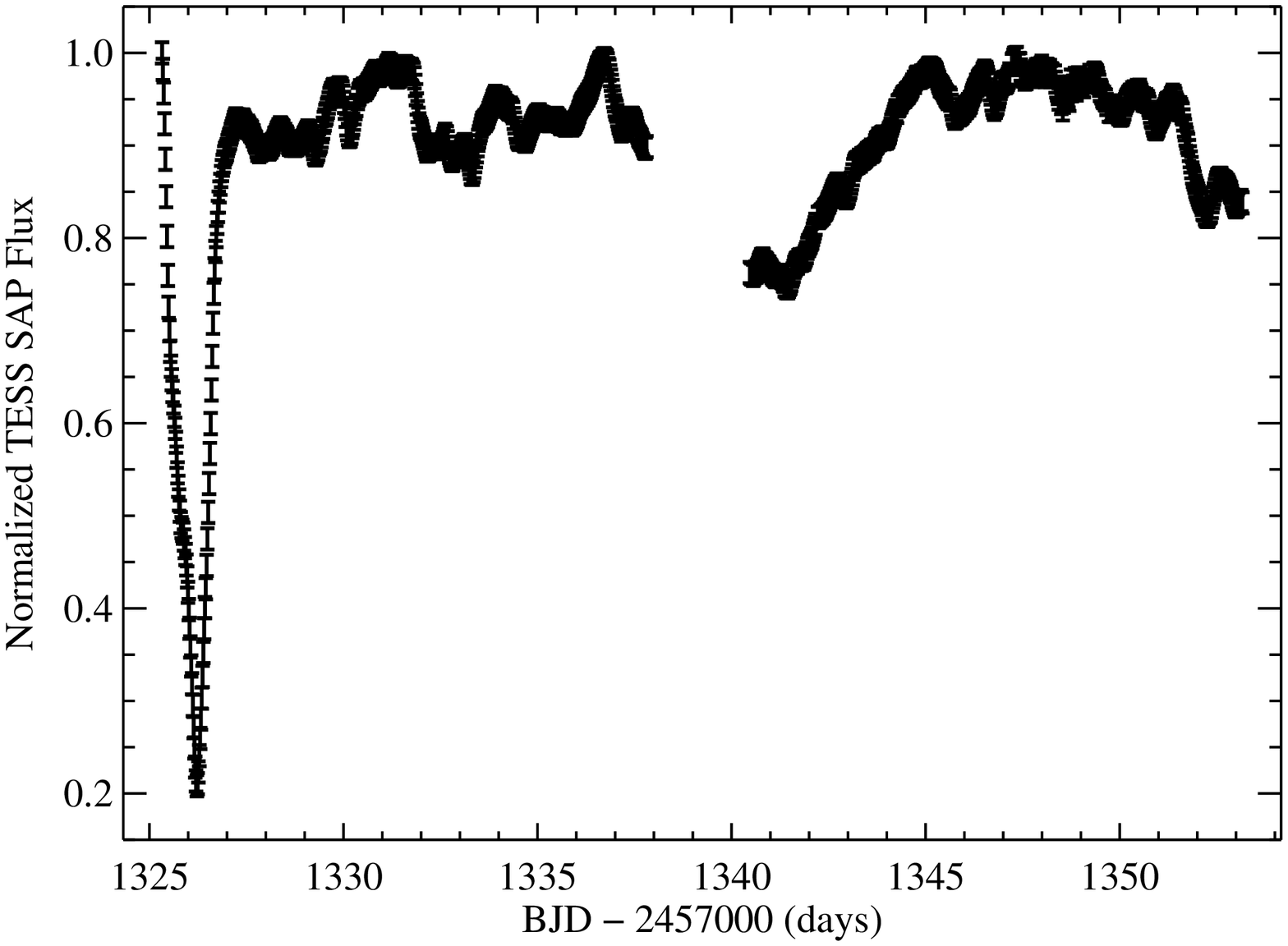}
 \end{minipage}
 \\*
 \begin{minipage}{155mm}
  \includegraphics[width=155mm]{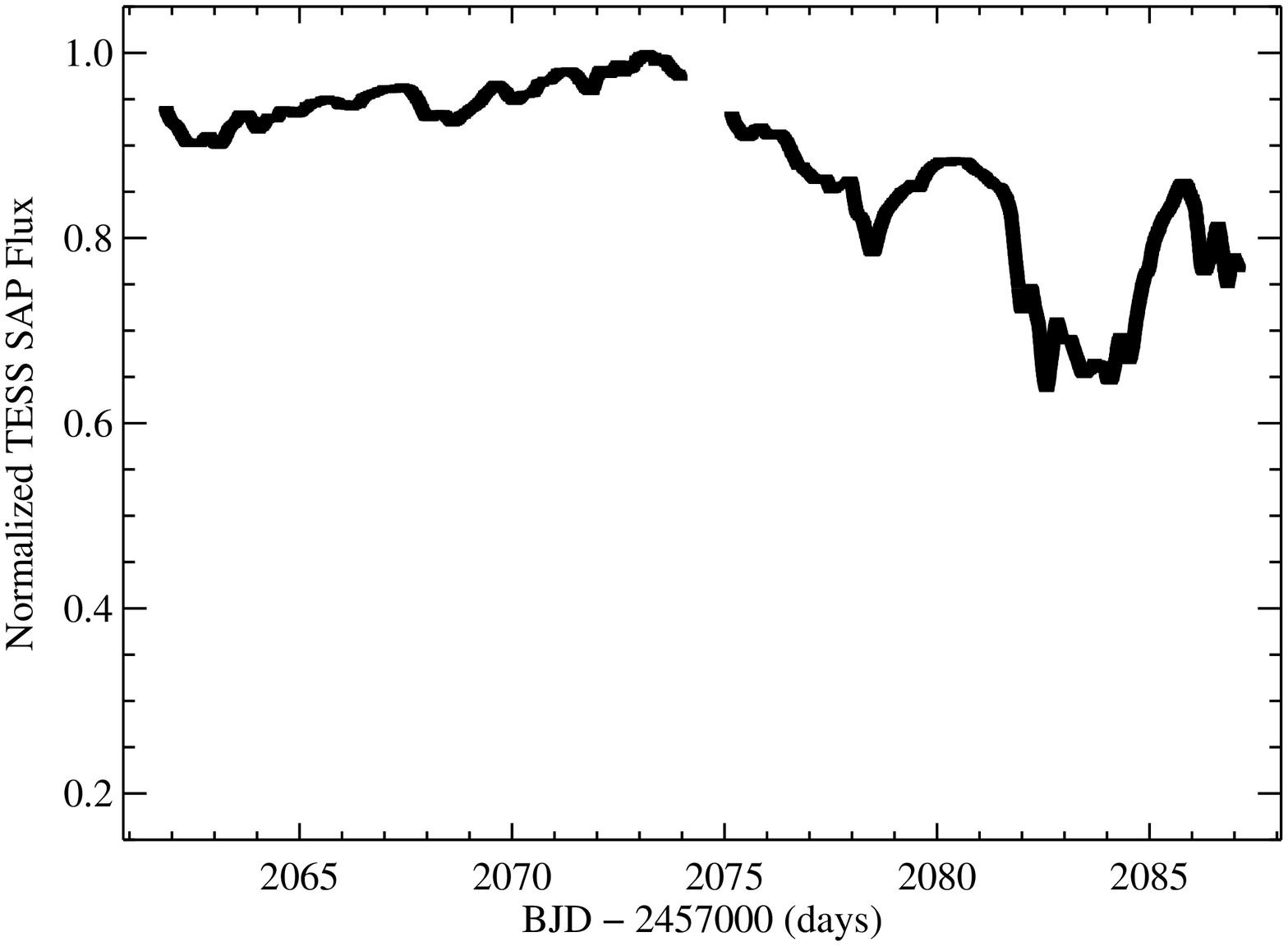}
 \end{minipage}
 \caption{\label{figtess} \large{ {\it TESS} lightcurves for TYC\,8830~410~1;
               flux is normalized to the maximum value seen in each individual panel. A
               deep dip and stochastic variability at the $\sim$20\% level are apparent.
               Each lightcurve is normalized to unity at its respective maximum value.
               {\it Top:} {\it TESS} Sector~1 FFI lightcurve.
               {\it  Bottom:} {\it TESS} Sector~28 two-minute cadence lightcurve.}}
 \end{figure}
 
  \begin{figure}
 \centering
 \begin{minipage}{160mm}
  \includegraphics[width=170mm]{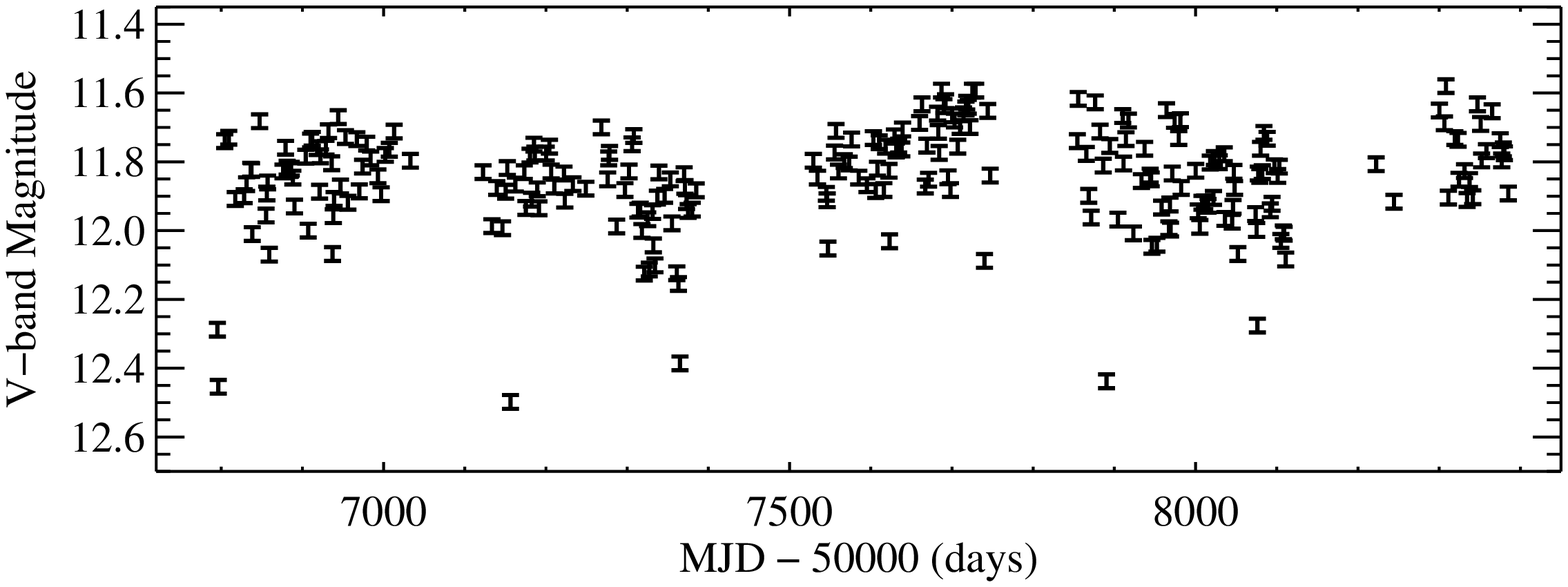}
 \end{minipage}
 \\*
 \begin{minipage}{160mm}
  \includegraphics[width=170mm]{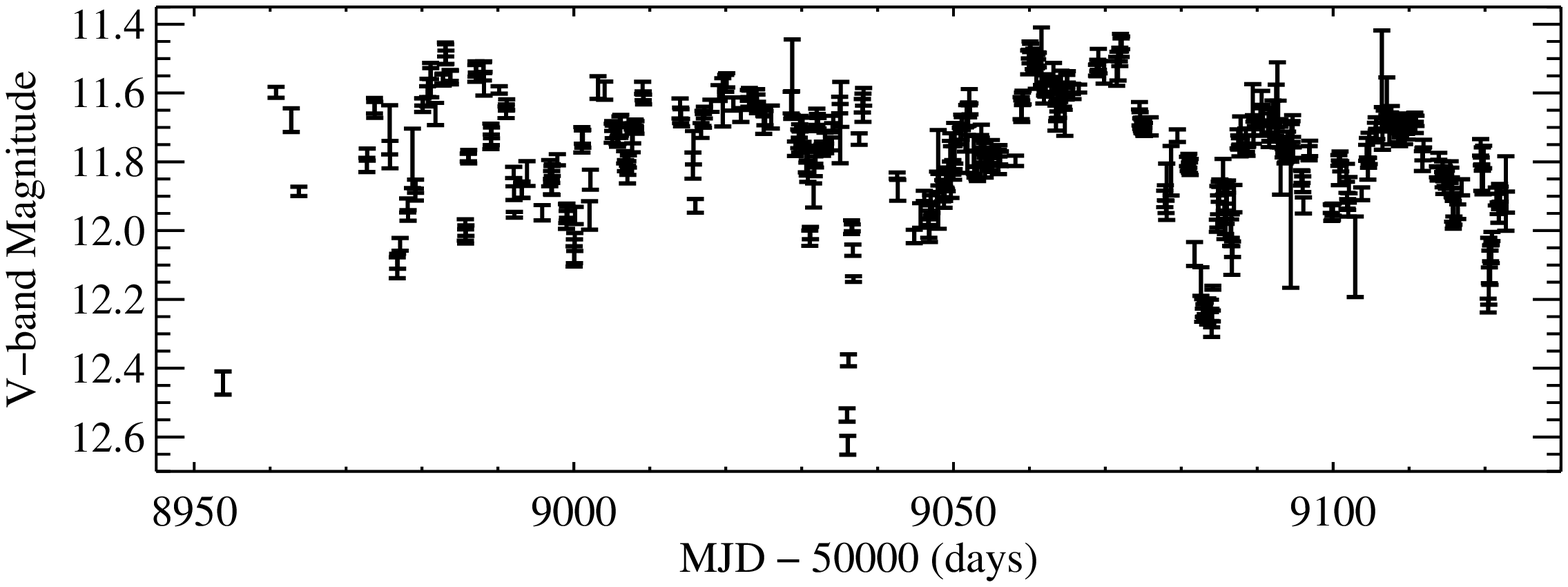}
 \end{minipage}
 \caption{\label{figasassnlco} \large{Ground-based optical lightcurves for TYC\,8830~410~1. Several
               $\approx$1~magnitude deep dips and stochastic variability at the $\sim$0.5~magnitude 
               level are apparent.
               {\it Top:} ASAS-SN lightcurve.
               {\it  Bottom:} LCOGT lightcurve.}}
 \end{figure}

\begin{figure}
 \centering
  \includegraphics[width=100mm]{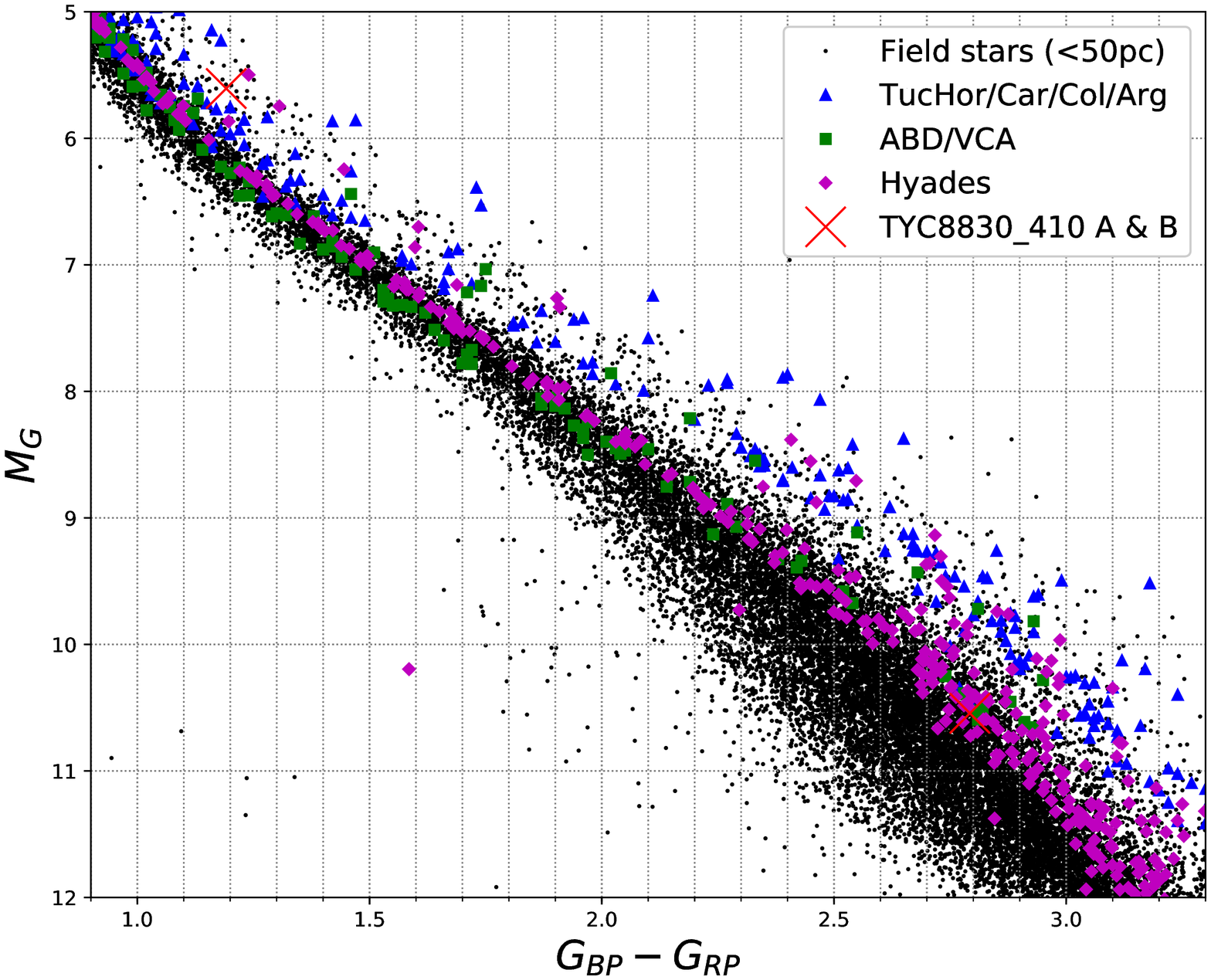}
  \\*
 \begin{minipage}{80mm}
  \includegraphics[width=80mm]{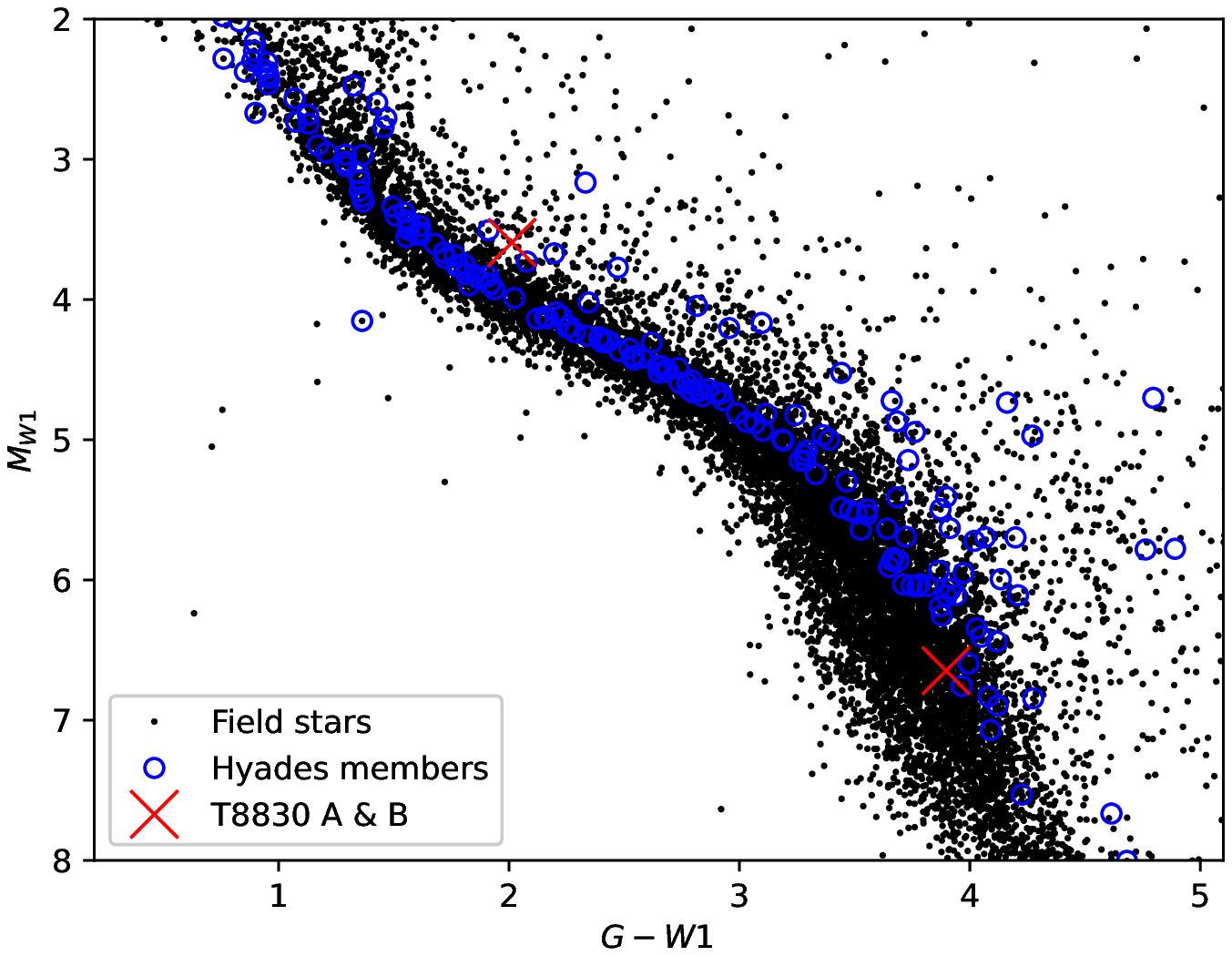}
 \end{minipage}
 \begin{minipage}{80mm}
  \includegraphics[width=80mm]{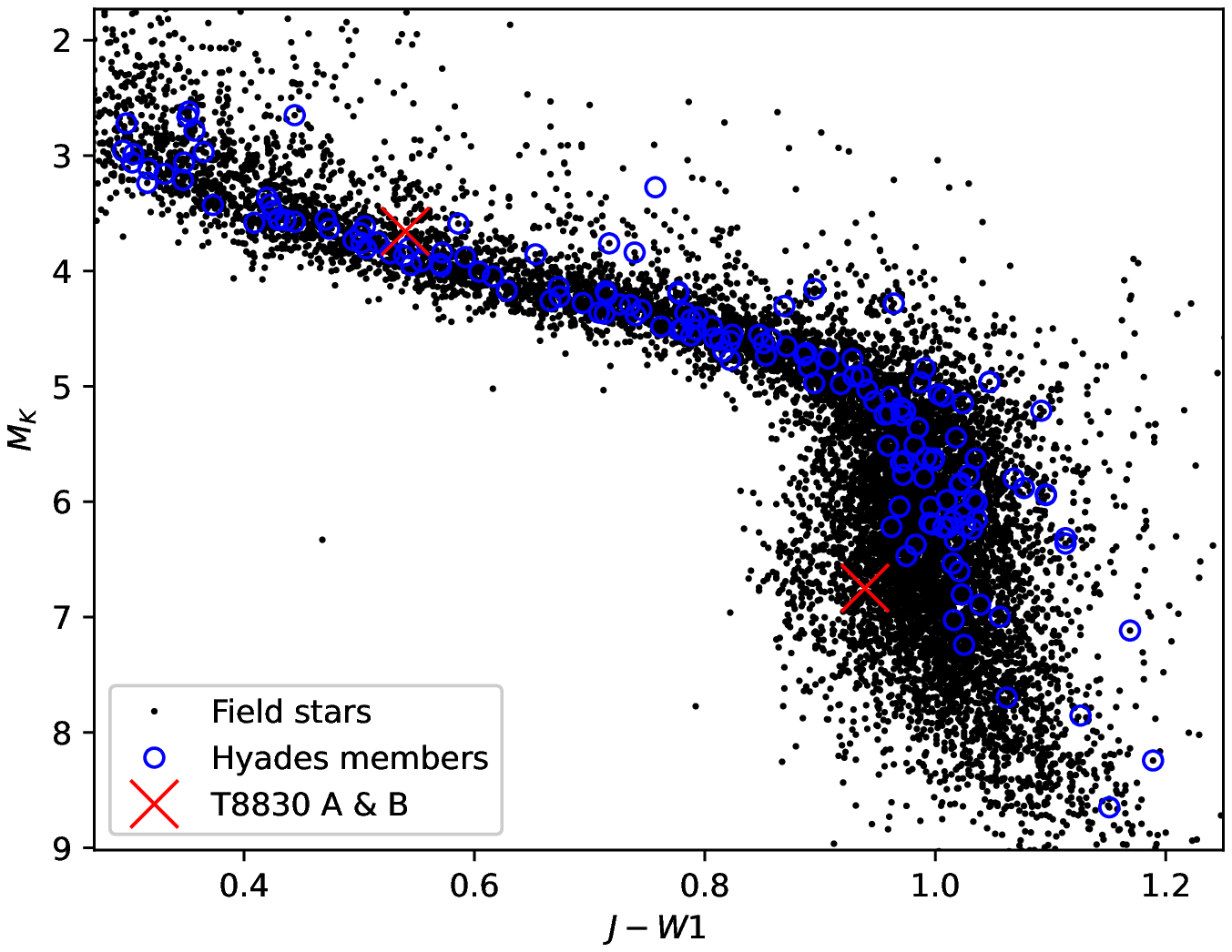}
 \end{minipage}
 \caption{\label{figiso} \large{Color-absolute magnitude diagrams constructed from {\it Gaia} EDR3
              photometry and parallax measurements and infrared photometry. 
              {\it Top panel:} Tucana-Horologium (TucHor), Carina (Car),
              Columba (Col), and Argus (Arg) associations have ages of $\sim$30\,Myr. AB Dor (ABD)
              and Volans-Carina (VCA) have ages of $\sim$100\,Myr
              (ages from \citealt{torres08}; \citealt{gagne18b}, and references therein).
              {\it All panels:} The Hyades age is $\approx$680\,Myr
              (e.g., \citealt{gossage18} and references therein). 
              TYC 8830 410 1 (TYC8830\_410 A) is either $\approx$0.2 magnitudes
              too red or too luminous (or possibly some combination of both)
              compared to single stars of any age (unresolved, similar
              brightness binary systems in the Hyades are seen above the main Hyades-locus
              with comparable positions as TYC 8830 410 1).
              Given the presence of transiting dust clumps, we assume reddening is responsible
              for its offset position. 2MASS\,J2301$-$5858 (TYC8830\_410 B) lies within the
              locus of Hyades mid-M dwarfs in the optical color-absolute magnitude diagram,
              although it may be compatible with $\sim$100\,Myr
              old stars. In the infrared color-absolute magnitude diagrams it is more consistent with
              Hyades and field
              mid-M type stars within 50\,pc
              of the Sun, which are thought to have ages on the order of gigayears.} }
\end{figure}
 

\clearpage
\begin{deluxetable}{lcc}
\tabletypesize{\large}
\tablecolumns{3}
\tablewidth{0pt}
\tablecaption{\large Stellar Parameters\label{tabstarpars}}
\tablehead{
 \colhead{Parameter} &
 \colhead{TYC\,8830~410~1} &
 \colhead{2MASS\,J23011901$-$5858262} 
}
\startdata
R.A.\ (J2016) & 23 01 12.637 & 23 01 19.017 \\
Decl.\ (J2016) & $-$58 58 21.74 & $-$58 58 26.07 \\
G$_{\rm mag}$ & 11.60 & 16.54 \\
G$_{\rm BP}$$-$G$_{\rm RP}$ & 1.19 & 2.79 \\
Sp.\ Type & G9\,V & M4\,Ve  \\ 
T$_{\rm star}$ (K) & 5300 & 3300 \\
R$_{\rm star}$ (R$_{\odot}$) & 0.96 & 0.35 \\
L$_{\rm star}$ (L$_{\odot}$) & 0.65 & 0.013 \\
pmRA (mas yr$^{-1}$)& +27.01$\pm$0.01 & +29.35$\pm$0.04 \\
pmDE (mas yr$^{-1}$) & $-$14.24$\pm$0.01 & $-$14.71$\pm$0.04 \\
RV (km s$^{-1}$) & +9.9$\pm$0.1 & +9.20$\pm$0.84 \\
Parallax (mas) & 6.30$\pm$0.01 & 6.35$\pm$0.05 \\
Distance (pc) & 158.7$\pm$0.3 & 157.5$\pm$1.2 \\
UVW (km s$^{-1}$) & \multicolumn{2}{c}{$-$11.4$\pm$0.1, $-$19.1$\pm$0.1, $-$11.5$\pm$0.1} \\
\enddata
\tablecomments{
{R.A.\ and Decl., 
G$_{\rm mag}$, 
G$_{\rm BP}$$-$G$_{\rm RP}$ color,
proper motions (pmRA and pmDE), and parallax (and hence distance)
values are from {\it Gaia} DR2 and EDR3 \citep{gaia18,gaia20}.
Spectral types are from analysis presented in Section \ref{seccomp}.
T$_{\rm star}$ and R$_{\rm star}$ are extracted from atmospheric model fits to broadband
photomteric measurements (e.g., Figure \ref{fig8830sed}). 
L$_{\rm star}$ is then calculated using L$=$4$\pi$R$^{2}$$\sigma$$_{\rm SB}$T$^{4}$.
Radial velocities are measured from our high- and medium-resolution echelle spectra.
UVW space motions are reported for the primary (independent UVW space motion values for
the companion are consistent but less precise) and are relative 
to the Sun such that positive U is towards the Galactic center, positive V is in the direction of Galactic 
rotation, and positive W is toward the north Galactic pole \citep{johnson87}.
}
}
\end{deluxetable}

\clearpage
\begin{deluxetable}{cllc}
\tabletypesize{\large}
\tablecolumns{4}
\tablewidth{0pt}
\tablecaption{\large TYC\,8830~410~1 Radial Velocity Measurements \label{tabrvs}}
\tablehead{
 \colhead{Instrument} &
 \colhead{UT Epoch} &
 \colhead{Julian Date} &
 \colhead{RV} \\
 \colhead{} &
 \colhead{} &
 \colhead{} &
 \colhead{(km\,s$^{-1}$)}
}
\startdata
RAVE   &  2009-11-10 & 2455145.905162   &   7.0$\pm$1.8 \\
{\it Gaia} DR2  & 2015                     &  2457205.5            &    9.35$\pm$0.77 \\
FEROS  &  2019-08-22 & 2458717.684903  & 10.012$\pm$0.014 \\
FEROS  &  2019-08-25 & 2458720.843824  &  9.991$\pm$0.020  \\ 
FEROS  &  2019-08-27 & 2458722.617453  &  9.787$\pm$0.011   \\
FEROS  &  2019-08-28 & 2458723.702129  &  9.834$\pm$0.009  \\
FEROS  &  2019-08-29 & 2458724.675237  &  9.875$\pm$0.010 \\
FEROS &   2019-11-13 & 2458800.576946  &  9.794$\pm$0.013 \\
FEROS &   2019-11-18 & 2458805.589559  &  9.854$\pm$0.010 \\
\enddata
\tablecomments{The {\it Gaia} DR2 measurement is a median of all epochs regardless of whether or not
any variability is present; it is not yet possible to access individual epoch measurements.}
\end{deluxetable}

\clearpage
\begin{deluxetable}{llrc}
\tabletypesize{\large}
\tablecolumns{4}
\tablewidth{0pt}
\tablecaption{\large TYC\,8830~410~1 Broadband Photometry \label{tabphot}}
\tablehead{
 \colhead{Filter Name} &
 \colhead{$\lambda$} &
 \colhead{Magnitude} &
 \colhead{Flux} \\
 \colhead{} &
 \colhead{($\mu$m)} &
 \colhead{} &
 \colhead{(mJy)}
}
\startdata
$B$ & 0.43    & 12.85$\pm$0.25       & 30$\pm$7 \\ 
$V$ & 0.55    & 11.87$\pm$0.05   & 67.5$\pm$0.6 \\ 
$R$ & 0.70  & 11.34$\pm$0.05     &  85.9$\pm$0.8 \\ 
$I$  & 0.778  & 10.83$\pm$0.01      & 116$\pm$1 \\ 
$J$ & 1.235     & 10.13$\pm$0.02   & 142$\pm$3 \\ 
$H$ & 1.662    &  9.77$\pm$0.03     & 127$\pm$3 \\ 
$K_{\rm s}$ & 2.159 &  9.65$\pm$0.02 & 92$\pm$2 \\ 
$W1$ & 3.35  &  9.57$\pm$0.02       & 45$\pm$1 \\
$W2$ & 4.60  &  9.53$\pm$0.02       & 26.3$\pm$0.5 \\
VISIR PAH1 & 8.59 & $-$                  & 23$\pm$2 \\ 
$W3$ & 11.56 &  7.43$\pm$0.02       & 31.0$\pm$0.5 \\
$W4$ & 22.09 &  6.51$\pm$0.06      & 21$\pm$1 \\
\enddata
\tablecomments{{\it BVR} are in the Johnson-Cousins system.
$B$-band data are from AAVSO \citep{henden15}.
$V$- and $R$-band data are converted from {\it Gaia} EDR3 G$_{\rm mag}$ and
G$_{\rm BP}$$-$G$_{\rm RP}$
color\footnote{\url{https://gea.esac.esa.int/archive/documentation/GDR2/Data_processing/chap_cu5pho/sec_cu5pho_calibr/ssec_cu5pho_PhotTransf.html}}.
$I$-band data are from DENIS \citep{epchtein99,fouque00},
$JHK$$_{\rm s}$ are from 2MASS \citep{cutri03,skrutskie06}.
VISIR PAH1-filter data are described in Section \ref{secvisir}.
$W1W2W3W4$ are from {\it WISE} \citep{cutri12} and are not color corrected.}
\end{deluxetable}

\thispagestyle{gaiacolor}
\clearpage

\appendix

\section{LCOGT V-band Photometry for TYC\,8830~410~1}

Epoch V-band magnitudes are presented for TYC\,8830~410~1 
as measured with the LCOGT and described in Section \ref{seclcogt}.
There are 484 measurements presented.

\begin{deluxetable}{ccc}
\tabletypesize{\large}
\tablecolumns{3}
\tablewidth{0pt}
\tablecaption{\large TYC\,8830~410~1 LCOGT V-band Photometric Monitoring \label{tabappA}}
\tablehead{
 \colhead{Modified Julian Date} &
 \colhead{V$_{\rm mag}$} &
 \colhead{V$_{\rm mag}$ error}
}
\startdata
 58953.79529 & 12.44 & 0.03 \\
58960.78542 & 11.59 & 0.01 \\
58962.78541 & 11.67 & 0.03 \\
58963.78535 & 11.88 & 0.01 \\
58972.74410 & 11.80 & 0.02 \\
58972.78802 & 11.77 & 0.01 \\
58973.74359 & 11.63 & 0.02 \\
58973.78536 & 11.64 & 0.02 \\
58975.74365 & 11.70 & 0.07 \\
58975.78527 & 11.77 & 0.04 \\
58976.74606 & 12.10 & 0.03 \\
58976.78527 & 12.08 & 0.02 \\
58977.11856 & 12.04 & 0.01 \\
58978.11856 & 11.92 & 0.02 \\
58978.16576 & 11.95 & 0.01 \\
58978.74372 & 11.79 & 0.08 \\
58979.12868 & 11.89 & 0.01 \\
58979.16023 & 11.87 & 0.01 \\
58980.08512 & 11.63 & 0.01 \\
58980.11906 & 11.62 & 0.01 \\
58980.74374 & 11.59 & 0.01 \\
58980.78534 & 11.59 & 0.04 \\
58981.12450 & 11.53 & 0.02 \\
58981.16026 & 11.57 & 0.04 \\
58981.78528 & 11.66 & 0.03 \\
58982.74389 & 11.55 & 0.01 \\
58982.78544 & 11.56 & 0.01 \\
58983.07718 & 11.48 & 0.03 \\
58983.11857 & 11.47 & 0.01 \\
58983.16030 & 11.49 & 0.01 \\
58983.71338 & 11.54 & 0.01 \\
58983.74353 & 11.55 & 0.01 \\
58985.70794 & 11.99 & 0.03 \\
58985.74361 & 12.01 & 0.02 \\
58985.78535 & 12.00 & 0.01 \\
58986.08382 & 11.78 & 0.01 \\
58986.11858 & 11.78 & 0.01 \\
58986.16020 & 11.78 & 0.01 \\
58987.07881 & 11.54 & 0.02 \\
58987.11858 & 11.53 & 0.02 \\
58987.16201 & 11.52 & 0.01 \\
58988.07689 & 11.53 & 0.02 \\
58988.11859 & 11.52 & 0.01 \\
58988.16185 & 11.57 & 0.03 \\
58989.07814 & 11.72 & 0.03 \\
58989.11857 & 11.73 & 0.01 \\
58989.16130 & 11.71 & 0.01 \\
58990.17046 & 11.59 & 0.01 \\
58991.07937 & 11.65 & 0.01 \\
58991.11855 & 11.64 & 0.01 \\
58991.16073 & 11.63 & 0.01 \\
58992.07690 & 11.83 & 0.01 \\
58992.11854 & 11.89 & 0.01 \\
58992.16075 & 11.95 & 0.01 \\
58993.11854 & 11.88 & 0.02 \\
58993.16262 & 11.88 & 0.02 \\
58993.82708 & 11.83 & 0.03 \\
58995.82694 & 11.94 & 0.02 \\
58996.82693 & 11.83 & 0.03 \\
58997.04211 & 11.87 & 0.02 \\
58997.07688 & 11.84 & 0.01 \\
58997.11885 & 11.85 & 0.04 \\
58997.16221 & 11.82 & 0.01 \\
58997.82710 & 11.79 & 0.01 \\
58999.03521 & 11.95 & 0.03 \\
58999.07789 & 11.95 & 0.02 \\
58999.11857 & 11.95 & 0.01 \\
58999.16017 & 11.94 & 0.01 \\
59000.03521 & 12.07 & 0.02 \\
59000.07688 & 12.06 & 0.03 \\
59000.11858 & 12.03 & 0.02 \\
59000.16760 & 11.95 & 0.02 \\
59001.05897 & 11.73 & 0.03 \\
59001.12490 & 11.72 & 0.01 \\
59001.18373 & 11.73 & 0.02 \\
59002.06012 & 11.95 & 0.04 \\
59002.18790 & 11.85 & 0.02 \\
59003.18448 & 11.58 & 0.03 \\
59004.05878 & 11.59 & 0.02 \\
59005.05890 & 11.71 & 0.02 \\
59005.12125 & 11.69 & 0.01 \\
59005.18374 & 11.71 & 0.01 \\
59005.68382 & 11.72 & 0.02 \\
59005.74632 & 11.74 & 0.01 \\
59005.80880 & 11.73 & 0.01 \\
59006.05890 & 11.69 & 0.03 \\
59006.12128 & 11.70 & 0.01 \\
59006.18375 & 11.67 & 0.01 \\
59006.68398 & 11.78 & 0.01 \\
59006.80879 & 11.81 & 0.01 \\
59007.05895 & 11.83 & 0.02 \\
59007.12128 & 11.81 & 0.02 \\
59007.18392 & 11.79 & 0.01 \\
59007.68392 & 11.73 & 0.03 \\
59007.74630 & 11.72 & 0.02 \\
59007.81043 & 11.70 & 0.01 \\
59008.05893 & 11.68 & 0.01 \\
59008.12117 & 11.70 & 0.01 \\
59008.18367 & 11.70 & 0.01 \\
59009.05877 & 11.58 & 0.01 \\
59009.12116 & 11.60 & 0.01 \\
59009.18367 & 11.61 & 0.01 \\
59013.99631 & 11.64 & 0.02 \\
59014.05878 & 11.66 & 0.02 \\
59014.12164 & 11.66 & 0.01 \\
59014.18390 & 11.68 & 0.01 \\
59015.68377 & 11.81 & 0.03 \\
59015.74650 & 11.75 & 0.04 \\
59015.99626 & 11.92 & 0.01 \\
59016.74629 & 11.70 & 0.02 \\
59016.99694 & 11.67 & 0.02 \\
59017.06013 & 11.68 & 0.02 \\
59017.18792 & 11.65 & 0.01 \\
59018.05877 & 11.63 & 0.01 \\
59019.12656 & 11.60 & 0.02 \\
59019.62133 & 11.62 & 0.07 \\
59020.00060 & 11.57 & 0.02 \\
59020.12772 & 11.56 & 0.02 \\
59020.99732 & 11.63 & 0.01 \\
59021.99627 & 11.66 & 0.02 \\
59022.99631 & 11.60 & 0.01 \\
59023.06086 & 11.62 & 0.01 \\
59023.12131 & 11.60 & 0.01 \\
59023.18398 & 11.62 & 0.02 \\
59023.99634 & 11.61 & 0.02 \\
59024.05873 & 11.60 & 0.02 \\
59024.12125 & 11.62 & 0.02 \\
59024.18388 & 11.64 & 0.01 \\
59024.99631 & 11.69 & 0.02 \\
59025.05892 & 11.65 & 0.01 \\
59025.12127 & 11.65 & 0.01 \\
59025.18378 & 11.67 & 0.02 \\
59025.99821 & 11.67 & 0.03 \\
59028.62132 & 11.62 & 0.03 \\
59028.68395 & 11.63 & 0.04 \\
59028.74628 & 11.55 & 0.10 \\
59029.18468 & 11.76 & 0.02 \\
59029.62270 & 11.73 & 0.02 \\
59029.68379 & 11.71 & 0.01 \\
59029.74629 & 11.73 & 0.04 \\
59029.80878 & 11.68 & 0.02 \\
59030.12128 & 11.68 & 0.01 \\
59030.18374 & 11.72 & 0.01 \\
59030.62134 & 11.79 & 0.03 \\
59030.68389 & 11.78 & 0.01 \\
59030.74629 & 11.81 & 0.02 \\
59030.80880 & 11.81 & 0.04 \\
59031.07689 & 12.02 & 0.02 \\
59031.16112 & 12.00 & 0.01 \\
59031.58231 & 11.89 & 0.03 \\
59031.66028 & 11.82 & 0.01 \\
59031.74362 & 11.79 & 0.01 \\
59031.82697 & 11.79 & 0.02 \\
59031.99353 & 11.66 & 0.02 \\
59032.07687 & 11.68 & 0.02 \\
59032.16113 & 11.68 & 0.01 \\
59032.66029 & 11.75 & 0.01 \\
59032.74364 & 11.76 & 0.02 \\
59033.07688 & 11.73 & 0.02 \\
59033.16022 & 11.76 & 0.01 \\
59033.99352 & 11.74 & 0.01 \\
59034.07837 & 11.70 & 0.01 \\
59034.16330 & 11.68 & 0.02 \\
59035.00199 & 11.63 & 0.02 \\
59035.07710 & 11.71 & 0.08 \\
59035.16195 & 11.63 & 0.06 \\
59035.99360 & 12.53 & 0.01 \\
59036.08050 & 12.62 & 0.02 \\
59036.16376 & 12.37 & 0.01 \\
59036.57691 & 11.99 & 0.01 \\
59036.66033 & 11.99 & 0.01 \\
59036.74364 & 12.05 & 0.01 \\
59036.82701 & 12.14 & 0.01 \\
59037.57694 & 11.73 & 0.01 \\
59037.99355 & 11.63 & 0.01 \\
59038.07688 & 11.64 & 0.04 \\
59038.16021 & 11.61 & 0.02 \\
59042.57700 & 11.84 & 0.01 \\
59042.66029 & 11.88 & 0.03 \\
59044.83058 & 12.01 & 0.01 \\
59045.66428 & 11.95 & 0.03 \\
59046.16020 & 11.89 & 0.01 \\
59046.67066 & 11.96 & 0.02 \\
59046.74698 & 11.99 & 0.02 \\
59046.82945 & 12.01 & 0.02 \\
59046.91588 & 11.94 & 0.01 \\
59046.99356 & 11.90 & 0.01 \\
59047.07689 & 11.93 & 0.01 \\
59047.16478 & 11.93 & 0.01 \\
59047.91151 & 11.85 & 0.02 \\
59047.99357 & 11.85 & 0.14 \\
59048.08371 & 11.85 & 0.01 \\
59048.57760 & 11.88 & 0.03 \\
59048.66029 & 11.87 & 0.02 \\
59048.74357 & 11.89 & 0.04 \\
59048.82694 & 11.89 & 0.02 \\
59048.99355 & 11.84 & 0.02 \\
59049.08345 & 11.85 & 0.02 \\
59049.16663 & 11.85 & 0.03 \\
59049.57810 & 11.76 & 0.02 \\
59049.66030 & 11.81 & 0.09 \\
59049.74361 & 11.78 & 0.03 \\
59049.91166 & 11.81 & 0.02 \\
59049.99357 & 11.84 & 0.01 \\
59050.07689 & 11.78 & 0.03 \\
59050.16019 & 11.74 & 0.05 \\
59050.91024 & 11.72 & 0.02 \\
59050.99571 & 11.71 & 0.02 \\
59051.07690 & 11.68 & 0.01 \\
59051.16020 & 11.68 & 0.01 \\
59051.57698 & 11.73 & 0.03 \\
59051.74434 & 11.66 & 0.03 \\
59051.82694 & 11.77 & 0.05 \\
59051.92056 & 11.66 & 0.03 \\
59051.99420 & 11.64 & 0.02 \\
59052.07697 & 11.61 & 0.02 \\
59052.16019 & 11.64 & 0.01 \\
59052.66294 & 11.76 & 0.04 \\
59052.74359 & 11.80 & 0.03 \\
59052.82695 & 11.83 & 0.01 \\
59053.07817 & 11.83 & 0.01 \\
59053.16017 & 11.83 & 0.02 \\
59053.57693 & 11.76 & 0.04 \\
59053.66031 & 11.71 & 0.01 \\
59053.74359 & 11.74 & 0.01 \\
59053.82703 & 11.74 & 0.01 \\
59054.07684 & 11.76 & 0.01 \\
59054.16018 & 11.81 & 0.02 \\
59054.57695 & 11.79 & 0.02 \\
59054.91240 & 11.78 & 0.01 \\
59054.99355 & 11.76 & 0.02 \\
59055.07697 & 11.80 & 0.02 \\
59055.16047 & 11.77 & 0.01 \\
59055.91177 & 11.80 & 0.03 \\
59055.99720 & 11.76 & 0.01 \\
59056.07688 & 11.78 & 0.01 \\
59056.16033 & 11.79 & 0.02 \\
59058.16020 & 11.79 & 0.01 \\
59058.91347 & 11.64 & 0.03 \\
59058.99365 & 11.65 & 0.03 \\
59059.07698 & 11.61 & 0.01 \\
59059.16018 & 11.61 & 0.01 \\
59059.91025 & 11.51 & 0.03 \\
59059.99357 & 11.48 & 0.01 \\
59060.07684 & 11.48 & 0.03 \\
59060.16018 & 11.46 & 0.01 \\
59060.50315 & 11.50 & 0.02 \\
59060.58376 & 11.50 & 0.03 \\
59060.91019 & 11.55 & 0.03 \\
59060.99354 & 11.50 & 0.01 \\
59061.07686 & 11.49 & 0.01 \\
59061.16020 & 11.50 & 0.02 \\
59061.57696 & 11.49 & 0.08 \\
59061.91036 & 11.60 & 0.03 \\
59062.00081 & 11.57 & 0.02 \\
59062.07342 & 11.58 & 0.01 \\
59062.15665 & 11.57 & 0.03 \\
59062.90681 & 11.57 & 0.02 \\
59062.99027 & 11.57 & 0.01 \\
59063.07328 & 11.56 & 0.01 \\
59063.16158 & 11.53 & 0.02 \\
59063.49499 & 11.67 & 0.03 \\
59063.58854 & 11.65 & 0.02 \\
59063.65676 & 11.64 & 0.02 \\
59063.74006 & 11.62 & 0.03 \\
59063.82350 & 11.64 & 0.02 \\
59063.90673 & 11.60 & 0.02 \\
59063.99009 & 11.60 & 0.02 \\
59064.07325 & 11.59 & 0.01 \\
59064.15665 & 11.55 & 0.01 \\
59064.49227 & 11.61 & 0.02 \\
59064.57350 & 11.62 & 0.02 \\
59064.65665 & 11.66 & 0.05 \\
59064.73999 & 11.59 & 0.01 \\
59064.82350 & 11.60 & 0.01 \\
59064.91028 & 11.56 & 0.01 \\
59064.99009 & 11.55 & 0.01 \\
59065.65673 & 11.58 & 0.01 \\
59065.73999 & 11.59 & 0.02 \\
59066.49010 & 11.58 & 0.01 \\
59068.84212 & 11.53 & 0.01 \\
59068.90690 & 11.53 & 0.01 \\
59068.99000 & 11.52 & 0.01 \\
59069.07342 & 11.50 & 0.03 \\
59069.15705 & 11.52 & 0.01 \\
59069.83952 & 11.54 & 0.03 \\
59071.49014 & 11.53 & 0.04 \\
59071.66239 & 11.52 & 0.04 \\
59071.83391 & 11.49 & 0.01 \\
59071.90670 & 11.49 & 0.02 \\
59071.99016 & 11.45 & 0.02 \\
59072.07331 & 11.46 & 0.02 \\
59072.15705 & 11.45 & 0.01 \\
59074.49009 & 11.63 & 0.01 \\
59074.57339 & 11.67 & 0.01 \\
59074.65672 & 11.68 & 0.01 \\
59074.74009 & 11.69 & 0.01 \\
59074.82573 & 11.69 & 0.01 \\
59074.90665 & 11.70 & 0.01 \\
59074.98999 & 11.69 & 0.02 \\
59075.07333 & 11.71 & 0.01 \\
59075.15673 & 11.70 & 0.01 \\
59075.90708 & 11.69 & 0.02 \\
59077.82356 & 11.90 & 0.02 \\
59077.90690 & 11.89 & 0.02 \\
59077.98999 & 11.93 & 0.01 \\
59078.07348 & 11.88 & 0.08 \\
59078.65793 & 11.82 & 0.07 \\
59079.49009 & 11.72 & 0.01 \\
59080.82348 & 11.79 & 0.01 \\
59080.90667 & 11.80 & 0.02 \\
59080.98998 & 11.79 & 0.02 \\
59081.07331 & 11.81 & 0.01 \\
59081.15675 & 11.81 & 0.02 \\
59081.74596 & 12.06 & 0.03 \\
59082.58603 & 12.14 & 0.04 \\
59082.82332 & 12.23 & 0.03 \\
59082.90667 & 12.23 & 0.01 \\
59082.98998 & 12.23 & 0.01 \\
59083.07330 & 12.23 & 0.02 \\
59083.15676 & 12.22 & 0.01 \\
59083.49088 & 12.23 & 0.03 \\
59083.82336 & 12.21 & 0.01 \\
59083.90665 & 12.24 & 0.01 \\
59083.99001 & 12.29 & 0.01 \\
59084.07336 & 12.24 & 0.01 \\
59084.15668 & 12.16 & 0.01 \\
59084.74004 & 11.98 & 0.01 \\
59084.82337 & 11.93 & 0.06 \\
59084.90662 & 11.92 & 0.02 \\
59084.98997 & 11.88 & 0.02 \\
59085.07331 & 11.86 & 0.01 \\
59085.49165 & 11.82 & 0.03 \\
59085.57494 & 11.87 & 0.01 \\
59085.66036 & 11.93 & 0.01 \\
59085.74382 & 11.98 & 0.04 \\
59085.82334 & 11.98 & 0.02 \\
59085.90664 & 11.92 & 0.05 \\
59085.98997 & 11.91 & 0.03 \\
59086.07331 & 11.87 & 0.01 \\
59086.49216 & 12.00 & 0.03 \\
59086.57541 & 11.98 & 0.03 \\
59086.65878 & 12.08 & 0.04 \\
59086.74202 & 12.05 & 0.02 \\
59086.92207 & 11.90 & 0.03 \\
59087.49020 & 11.73 & 0.02 \\
59087.57343 & 11.74 & 0.02 \\
59087.65901 & 11.74 & 0.01 \\
59087.74235 & 11.76 & 0.01 \\
59087.82315 & 11.71 & 0.04 \\
59087.90870 & 11.74 & 0.01 \\
59088.49016 & 11.73 & 0.03 \\
59088.57607 & 11.75 & 0.01 \\
59088.65941 & 11.77 & 0.01 \\
59088.74276 & 11.74 & 0.01 \\
59089.42410 & 11.63 & 0.06 \\
59089.49013 & 11.70 & 0.04 \\
59089.57661 & 11.70 & 0.01 \\
59089.65978 & 11.68 & 0.01 \\
59089.74359 & 11.68 & 0.01 \\
59090.42128 & 11.66 & 0.02 \\
59090.49002 & 11.66 & 0.02 \\
59090.57348 & 11.64 & 0.04 \\
59090.66012 & 11.66 & 0.02 \\
59090.74876 & 11.68 & 0.04 \\
59091.41853 & 11.69 & 0.01 \\
59091.49012 & 11.69 & 0.03 \\
59091.57334 & 11.70 & 0.05 \\
59091.66083 & 11.72 & 0.01 \\
59091.74007 & 11.71 & 0.02 \\
59092.07327 & 11.65 & 0.03 \\
59092.41580 & 11.70 & 0.01 \\
59092.49008 & 11.63 & 0.06 \\
59092.65669 & 11.61 & 0.10 \\
59092.74005 & 11.69 & 0.06 \\
59093.00037 & 11.68 & 0.01 \\
59093.07600 & 11.82 & 0.06 \\
59093.41311 & 11.77 & 0.01 \\
59093.49007 & 11.75 & 0.01 \\
59093.57344 & 11.76 & 0.02 \\
59093.65669 & 11.77 & 0.03 \\
59093.74066 & 11.76 & 0.03 \\
59093.99016 & 11.73 & 0.01 \\
59094.41169 & 11.98 & 0.18 \\
59094.49008 & 11.72 & 0.03 \\
59094.57356 & 11.71 & 0.03 \\
59094.65677 & 11.69 & 0.03 \\
59094.74014 & 11.77 & 0.01 \\
59095.82404 & 11.84 & 0.02 \\
59095.90681 & 11.84 & 0.01 \\
59095.98997 & 11.86 & 0.02 \\
59096.07683 & 11.92 & 0.02 \\
59096.82825 & 11.77 & 0.02 \\
59096.90670 & 11.76 & 0.02 \\
59099.75746 & 11.93 & 0.01 \\
59099.82332 & 11.94 & 0.02 \\
59099.90665 & 11.94 & 0.02 \\
59099.98996 & 11.93 & 0.01 \\
59100.75474 & 11.79 & 0.01 \\
59100.82334 & 11.83 & 0.03 \\
59100.91616 & 11.78 & 0.01 \\
59100.98998 & 11.81 & 0.01 \\
59101.75390 & 11.83 & 0.02 \\
59101.82336 & 11.91 & 0.02 \\
59101.90668 & 11.92 & 0.01 \\
59101.99004 & 11.93 & 0.02 \\
59102.07830 & 11.86 & 0.02 \\
59102.90641 & 12.07 & 0.11 \\
59103.74648 & 11.89 & 0.01 \\
59104.41184 & 11.77 & 0.02 \\
59104.50743 & 11.75 & 0.02 \\
59104.57339 & 11.82 & 0.03 \\
59104.65673 & 11.80 & 0.01 \\
59104.74020 & 11.76 & 0.04 \\
59105.57349 & 11.71 & 0.02 \\
59105.65672 & 11.69 & 0.01 \\
59105.74007 & 11.71 & 0.01 \\
59106.40806 & 11.56 & 0.14 \\
59106.49005 & 11.70 & 0.06 \\
59106.57337 & 11.69 & 0.01 \\
59106.74227 & 11.66 & 0.01 \\
59107.07339 & 11.60 & 0.05 \\
59107.40669 & 11.71 & 0.03 \\
59107.49002 & 11.68 & 0.02 \\
59107.57336 & 11.67 & 0.01 \\
59107.65678 & 11.65 & 0.01 \\
59108.40677 & 11.69 & 0.01 \\
59108.49005 & 11.68 & 0.02 \\
59108.57338 & 11.73 & 0.01 \\
59108.65674 & 11.69 & 0.01 \\
59108.74015 & 11.72 & 0.03 \\
59108.82348 & 11.71 & 0.01 \\
59109.41375 & 11.70 & 0.03 \\
59109.49002 & 11.67 & 0.01 \\
59109.58051 & 11.69 & 0.01 \\
59109.65677 & 11.69 & 0.02 \\
59109.73981 & 11.70 & 0.03 \\
59109.83319 & 11.69 & 0.01 \\
59109.90663 & 11.67 & 0.01 \\
59110.74702 & 11.67 & 0.01 \\
59110.82351 & 11.70 & 0.01 \\
59110.91291 & 11.68 & 0.01 \\
59111.74676 & 11.79 & 0.02 \\
59111.82333 & 11.77 & 0.01 \\
59111.90663 & 11.74 & 0.01 \\
59113.74000 & 11.82 & 0.01 \\
59113.82330 & 11.83 & 0.01 \\
59113.90664 & 11.78 & 0.01 \\
59114.40678 & 11.81 & 0.01 \\
59114.49004 & 11.85 & 0.01 \\
59114.57341 & 11.87 & 0.01 \\
59114.65677 & 11.83 & 0.02 \\
59114.74010 & 11.84 & 0.02 \\
59114.82331 & 11.85 & 0.01 \\
59114.90664 & 11.85 & 0.02 \\
59115.40673 & 11.83 & 0.03 \\
59115.49003 & 11.84 & 0.03 \\
59115.57338 & 11.86 & 0.02 \\
59115.65703 & 11.88 & 0.01 \\
59115.74001 & 11.93 & 0.03 \\
59115.82325 & 11.95 & 0.04 \\
59115.90664 & 11.97 & 0.01 \\
59116.41185 & 11.94 & 0.02 \\
59116.90670 & 11.87 & 0.02 \\
59119.42069 & 11.75 & 0.02 \\
59119.49857 & 11.79 & 0.01 \\
59119.57340 & 11.84 & 0.03 \\
59119.65697 & 11.79 & 0.03 \\
59119.74015 & 11.85 & 0.03 \\
59119.82336 & 11.78 & 0.03 \\
59120.41186 & 12.21 & 0.02 \\
59120.49116 & 12.11 & 0.09 \\
59120.57340 & 12.13 & 0.02 \\
59120.65957 & 12.10 & 0.04 \\
59120.74000 & 12.06 & 0.02 \\
59120.82332 & 12.06 & 0.02 \\
59120.90663 & 12.01 & 0.01 \\
59121.49022 & 11.90 & 0.02 \\
59121.74004 & 11.93 & 0.01 \\
59121.82338 & 11.93 & 0.04 \\
59121.90676 & 11.90 & 0.03 \\
59121.98995 & 11.89 & 0.02 \\
59122.73995 & 11.89 & 0.10 \\
59122.82330 & 11.91 & 0.03 \\
\enddata
\end{deluxetable}

\end{document}